%% file: seds_2.tex
\documentclass{aa}
\usepackage{txfonts}
\usepackage{graphicx}
\usepackage{rotating} 
\usepackage{psfig}
\usepackage{latexsym}
\usepackage{amssymb}
\usepackage{natbib}
\usepackage[innercaption]{sidecap}

\def\cm3{${\rm cm^{-3}}$}            
\def\mum{${\rm \, \mu m}$}                            
\def\kms{\hbox{\,km\,s$^{-1}$}}  

\newcommand{\Teff}{\ensuremath{T_\mathrm{eff}}}
\newcommand{\HI}{\ion{H}{i}}
\newcommand{\HII}{\ion{H}{ii}}
\newcommand{\HeI}{\ion{He}{i}}
\newcommand{\HeII}{\ion{He}{ii}}
\newcommand{\kNIII}{\ion{N}{iii}} 
\newcommand{\kCIV}{\ion{C}{iv}} 
\newcommand{\kCIII}{\ion{C}{iii}} 
\newcommand{\NII}{[\ion{N}{ii}]}
\newcommand{\NIII}{[\ion{N}{iii}]}
\newcommand{\ArII}{[\ion{Ar}{ii}]}
\newcommand{\ArIII}{[\ion{Ar}{iii}]}
\newcommand{\SIII}{[\ion{S}{iii}]}
\newcommand{\SIV}{[\ion{S}{iv}]}
\newcommand{\NeII}{[\ion{Ne}{ii}]}
\newcommand{\NeIII}{[\ion{Ne}{iii}]}

\def\neratio{[\ion{Ne}{iii}]/[\ion{Ne}{ii}]}
\def\arratio{[\ion{Ar}{iii}]/[\ion{Ar}{ii}]}
\def\sratio{[\ion{S}{iv}]/[\ion{S}{iii}]}

\begin{document}

   \title{Metallicity and the spectral energy distribution and spectral types of dwarf O-stars}

   \author{M.\,R. Mokiem\inst{1} \and
           N.\,L. Mart\'{\i}n-Hern\'{a}ndez\inst{2}\thanks{\emph{Present
                address:} Geneva Observatory, 1290 Sauverny,
                Switzerland} \and
           A. Lenorzer\inst{1} \and
           A. de Koter\inst{1} \and
           A.\,G.\,G.\,M. Tielens\inst{2,3}
          }

   \offprints{M.\,R. Mokiem (mokiem@science.uva.nl)}

   \institute{
     Astronomical Institute Anton Pannekoek, University of
     Amsterdam, Kruislaan 403, 1098 SJ Amsterdam, The Netherlands
     \and
     Kapteyn Institute, P.O. Box 800, 9700 AV Groningen, The Netherlands
     \and
     SRON, National Institute for Space Reasearch, P.O. Box 800,
     9700 AV Groningen, The Netherlands 
   }

   \date{Received 5 December 2002 / Accepted 25 December 2003}

\abstract{We present a systematic study of the effect of metallicity
on the stellar spectral energy distribution (SED) of O main sequence
(dwarf) stars, focussing on the hydrogen and helium ionizing continua,
and on the optical and near-IR lines used for spectral classification.
The spectra are based on non-LTE line blanketed atmosphere models with
stellar winds calculated using the {\sc cmfgen} code of
\cite{hillier98}. We draw the following conclusions. First, we find
that the total number of Lyman photons emitted is almost independent
of line blanketing effects and metallicity for a given effective
temperature. This is because the flux that is blocked by the forest of
metal lines at $\lambda < 600$ \AA\ is redistributed mainly within the
Lyman continuum. Second, the spectral type, as defined by the ratio of
the equivalent widths of \HeI~$\lambda$4471 and \HeII~$\lambda$4542,
is shown to depend noticeably on the microturbulent velocity in the
atmosphere, on metallicity and, within the luminosity class of
dwarfs, on gravity. Third, we confirm the decrease in \Teff\ for a
given spectral type due to the inclusion of line blanketing recently
found by e.g. \citet{martins02}. Finally, we find that the SED below
$\sim 450$ \AA\ is highly dependent on metallicity. This is reflected
in the behaviour of nebular fine-structure line ratios such as
\neratio\ 15.5/12.8 and \arratio\ 9.0/7.0 \mum. This dependence
complicates the use of these nebular ratios as diagnostic tools for
the effective temperature determination of the ionizing stars in \HII\
regions and for age dating of starburst regions in galaxies.

 }

\maketitle

\section{Introduction}

Massive stars are a dominant force in the evolution of the
interstellar medium of galaxies. The extreme ultraviolet photons of
massive stars can ionize surrounding atomic hydrogen gas. This ionized
gas is heated by the photo electrons and cooled through forbidden line
radiation of trace species such as oxygen, resulting in a temperature
of some $10^4$~K. Besides this energetic coupling through the
radiation field, massive stars can also influence their surroundings
dynamically. The high pressure of the ionized gas will drive shock
waves in their surroundings, sweeping up the ambient molecular
gas. Moreover, these stars also have strong stellar winds and explode
as type {\sc ii} supernovae, both of which provide kinetic energy to
the interstellar gas. Understanding the characteristics of massive
stars and their interaction with their environment is therefore a key
problem in astrophysics.

The stellar spectral energy distribution in the extreme ultraviolet
(EUV) controls the ionization structure of \HII~ regions. However,
because of the high opacity of neutral gas in the EUV, this wavelength
range cannot be observed directly. Spectral typing of stars is
generally done through optical
\citep[e.g.][]{conti71,conti77,mathys88} or near-IR features
\citep[e.g.][]{hanson96,meyer98,hanson98,lenorzer02}. Alternatively,
observed nebular ionization structures can be used to probe the
ionizing fluxes of the stars energizing the medium.
This technique is now widely used to determine the properties of stars
in the EUV and has even evolved to a more general astronomical tool,
particularly for the study of regions where individual stars cannot be
directly typed
\cite[e.g.][]{oey00,takahashi00,okamoto01,morisset:paperiii}.
Moreover, because the hardness of the EUV radiation field is a good
measure of the spectral type of the ionizing star and because later
spectral types live longer, the observed ionization structure can be
used as an age indicator of a starburst region
\citep[e.g.][]{crowther99,thornley00,spoon00}.

In recent years, some problems with the latter technique have
emerged. For example, for the well-studied \HII~ region G29.96$-$0.02,
the spectral type derived from direct observations of the stellar
lines in the near-IR \citep{watson97b,martin:G29} indicates a much
earlier spectral type than that obtained from the ionization structure
\citep{morisset:paperiii}. This seems to be a general problem and may
be related to the metallicity of the region. Indeed, there is a loose
correlation between the ionization structure - as measured by for
example the \neratio~15.5/12.8 \mum\ line ratio - and the metallicity
of \HII~ regions \cite[see][]{martin:metal}. Such a correlation may
reflect the effects of metallicity on line blanketing or on the
characteristics of the stellar wind.

Much theoretical effort has been dedicated to best describe the EUV
spectra of massive stars. This is a formidable task because these
stars have strong winds and extended atmospheres. This leads to strong
non-LTE effects in the formation of spectral lines. Over the last ten
years much progress has been made and current models include the
effects of tens of thousands of lines on the energy balance and
temperature structure of the stellar photosphere and wind. So far
relatively modest effort has been investigated in systematic studies
of the effects of metallicity on the stellar spectral energy
distribution. In particular, there is no good theoretical
understanding of the effects of metallicity on the ionizing fluxes of
massive stars or on the optical and near-IR spectral characteristics
used to type these stars.  Here, we study the influence of metallicity
on the spectral energy distribution of O stars and determine its
influence on the resulting ionization structure of \HII\ regions.

This paper is organized as follows. Sect.~\ref{sect:seds:models}
presents a set of main-sequence (dwarf) star models constructed using
the {\sc cmfgen} code by \cite{hillier98} and compares the predicted
EUV fluxes with those from other codes.
Sect.~\ref{sect:seds:description} presents a detailed analysis of the
variations of the EUV spectral appareance and ionizing fluxes with
effective temperature and metallicity.  Sect.~\ref{sect:seds:sptype}
investigates the influence of metallicity and other stellar parameters
on the optical and near-IR spectral calibration. In
Sect.~\ref{sect:seds:hii}, the ionizing structure of single star \HII\
regions is studied. Finally, Sect.~\ref{sect:seds:conclusions}
presents the conclusions of this study.

\section{Description of the models}
\label{sect:seds:models}

The models presented in this paper are constructed using the {\sc
cmfgen} program of \cite{hillier98}, to which we refer for a full
description of techniques. In short: {\sc cmfgen} iteratively solves
the equations of radiative transfer subject to the constraints of
statistical and radiative equilibrium, for an atmosphere with an
outflowing stellar wind. The ions included in the non-LTE calculation
are \ion{H}{i}, \ion{He}{i-ii}, \ion{C}{iv}, \ion{N}{iii-v},
\ion{O}{iv-vi}, \ion{Si}{iv} and \ion{Fe}{iv-vii}. The cumulative
effect of the iron lines, causing a line blanketing effect, is
self-consistently accounted for by employing a description of the
atomic models in terms of super levels. All and all, approximately
9\,000 transitions are accounted for.

We used this code to calculate a grid consisting of eight
main-sequence (dwarf) star models ranging in effective temperature
from $33\,000$~K up to $51\,000$~K. Basic stellar parameters, \Teff,
$R_\ast$ and $M_\ast$, using masses predicted by evolution theory, are
taken from the Vacca et al. (1996) calibration and are given in
Table~\ref{table:seds:grid}.
  
We have opted to describe the photospheric density structure using a
constant scaleheight
\begin{equation}
  H = \frac{kT_{\rm eff}}{\mu m_{\rm amu} g_{\rm eff}}~,
\end{equation}
where $\mu$ is the mean molecular weight in atomic mass units, $g_{\rm
eff}$ is the effective gravity at the stellar surface corrected for
radiation pressure by electron scattering and all other parameters
have their usual meaning. Near and beyond the sonic point, the density
is set by the velocity law through the equation of
mass-continuity. The connection between photosphere and wind is
smooth. The wind velocity structure is given by a standard $\beta$-law
adopting $\beta = 0.8$, which is representative for dwarf O stars
\citep{groenewegen89}. The terminal flow velocity $v_{\infty}$ follows
from a scaling with the escape velocity \citep{lamers95} and is also
listed in Table~\ref{table:seds:grid}. The mass-loss rates are from
predictions by \cite{vink00,vink01}. These authors show that for
O-type stars the mass loss scales with the metal content $Z$ as
\begin{equation}
  \dot{M} \propto Z^{0.85}~.
\nonumber
\end{equation}
\cite{leitherer92} showed that the final wind velocity also
depends on metallicity as $v_\infty \propto Z^{0.13}$.We do not
incorporate this dependence in the models.
The microturbulent velocity was set to 20~\kms\ for the atmospheric
structure calculations. For the calculation of the emergent spectrum
we assumed microturbulent velocities of 20, 10 and 5~\kms.

In order to investigate the spectral appearance as a function of
metallicity, we have calculated an additional set of models for our
models \#2 and \#5, in which we set the metal content to $Z$ = 2, 1/2,
1/5 and 1/10 $Z_{\sun}$. In conjunction with $Z$, we adjusted the
helium abundance $Y$ according to
\begin{equation}
  Y = Y_{\rm p} + \left(\frac{\Delta Y}{\Delta Z}\right) Z~,
\end{equation}
where $Y_{\rm p} = 0.24$ is the primordial helium abundance
\citep{audoze87} and $(\Delta Y/\Delta Z) = 3$ is an observed constant
\citep{pagel92}. To arrive at the appropriate metal abundances, we
simply scaled the solar values \citep{cox00}, which we give in
Table~\ref{table:seds:chemical} for reference. The mass-loss rates and
scaleheights, which also depend on $Z$ through the mean molecular
weight, of the additional models are given in
Table~\ref{table:seds:grid:z}.

\input{./table1.tex}

\subsection{The assumption of a constant scaleheight}
\label{sect:seds:scaleheight}

The assumption of a constant scaleheight could lead to an
overestimation of the effective gravity as derived from density
sensitive lines. To test the validity of this assumption we
recalculated model \#6 using an atmospheric structure calculated with
the {\sc isa-wind} code of \cite{dekoter93, dekoter97}. This provides
a more realistic density stratification as it accounts for the effects
of continuum radiation pressure on the atmosphere (see section
\ref{sect:seds:comparison}). For the comparison we focused on the
spectral signatures investigated in this paper, i.e.\ the EUV, the
optical \HeI~$\lambda$4471 and \HeII~$\lambda$4542 lines and the
photospheric lines in the near-IR K-band.

\input{./table2.tex}

\input{./table3.tex}

Close inspection of the predicted EUV spectra shows that the number of
ionizing photons below a certain wavelength, as well as the total
amount of ionizing photons in the hydrogen and helium continua, agree
within a few percent. The changes in the optical \HeI\ and \HeII\
lines are also negligible. A decrease of at most 3\% in equivalent
width is observed compared to the constant scaleheight model.

A fair comparison for the near-IR lines is more difficult, as many of
these lines are partly formed in the transition zone where the
photosphere connects to the stellar wind. In both the {\sc isa-wind}
as in the constant scaleheight model this connection is made in an ad
hoc fashion. Keeping this in mind, we find that the differences in
equivalent width between the two models is at most 4\% for the \HeI\
lines and at most 12\% for the \kCIV\ lines.

\subsection{Comparison with other model atmosphere codes}
\label{sect:seds:comparison}

\input{./table4.tex}

Before comparing ionizing fluxes predicted by our models with those
from other codes, we first briefly discuss the basic physics treated
in the model atmospheres used for this comparison.

We focus on a comparison of main sequence (dwarf) O star models
computed using the {\sc cmfgen} code (this study), the {\sc WM-basic}
code of \cite{pauldrach01} and the {\sc CoStar} models of
\cite{schaerer97}. The latter make use of the {\sc isa-wind} code of
\cite{dekoter93, dekoter97} to predict the stellar spectrum. A
schematic comparison of assumptions made in these three codes is given
in Table~\ref{table:seds:codes}. Note that a comparison of codes does
not address the question concerning the level of realism of the
spectra produced. Assessing this issue requires empirical testing by
analyzing \HII\ regions containing single hot stars of which {\em
ideally} the basic parameters have been determined from detailed
quantitative spectroscopy using the same models as those applied to
predict the ionizing fluxes. This approach has been followed by
\cite{crowther99}, who tested the {\sc cmfgen} and {\sc isa-wind}
models for a nebula containing a cool WN star and found similar
results except for the ionizing flux distributions below the \HeI\
edge at 504 \AA, which were quite different. The next best thing is to
study \HII\ regions containing single stars with well defined spectral
types and effective temperatures. \cite{schaerer00} reviews the
success of non-LTE model atmospheres in reproducing diagnostic line
ratios of nebula irradiated by O-type stars. The metallicity
dependence of the ionizing flux may be studied by modelling
extra-galactic \HII\ regions \citep[e.g.][]{bresolin99,oey00,
bresolin2002}.

A comparison of ionizing fluxes from different codes is
straightforward, however, interpreting the differences is --
unfortunately -- not. Modest discrepancies are expected to originate
from small differences in the adopted abundances and photosphere \&
wind parameters of the models one is comparing. For instance, a
difference in effective temperature of 1000~K yields a flux difference
in the Wien part of the spectrum of 10--15\%. This may seem a needless
`mistake' and one that is easily fixed. However, these complex models
are difficult to compute and it turns out to be less straightforward
than one may think to tune all parameters. Having said this, the main
sources of differences between codes likely is in the way in which
essential physical processes are treated. Notably {\em i)} the
treatment of the equation of motion,
{\em  ii)} the exact treatment of the non-LTE rate equations,
{\em iii)} the number of lines included (affecting the amount of line
blocking), and
{\em  iv)} the way line blanketing effects are taken into account.

In the {\sc cmfgen} models we present here, we have described the
photospheric density structure in a relatively simplistic way (see
above). The {\sc CoStar} models, using an atmospheric structure
calculated with the {\sc isa-wind} code, have a more sophisticated
density stratification. Here a grey LTE temperature structure and an
Euler equation taking into account continuum radiation pressure are
iterated until convergence, accounting for the feedback of metal line
blanketing on the Rosseland optical depth (see below). The wind is
described in a similar way as for the {\sc cmfgen} models. The {\sc
WM-basic} models have the most self-consistent description of the
density structure, as a final iteration is done, based on the non-LTE
temperature structure, in which both the continuum and line force is
included in the equation of motion. This yields the density structure
throughout the photosphere and wind, i.e. no {\em ad hoc} wind
velocity law is required.

The {\sc cmfgen} and {\sc WM-basic} models compute the non-LTE state
of the gas in a self-consistent way, using a super-level formalism. In
total these codes treat $\sim$ 9\,000 and $\sim$ 30\,000 transitions,
respectively. In this respect, the {\sc CoStar} models are relatively
simple as the statistical equilibrium equations are solved explicitly
for H, He, CNO and Si only, yielding $\sim$ 1\,000 transitions, while
the iron group elements, adding $\sim$ 100\,000 lines,
are treated in a modified nebular approximation. Moreover, the Sobolev
approximation is applied to describe line transfer.  Note that {\sc
cmfgen} treats $\sim$ 17\,000 individual lines in the emergent
spectrum, while {\sc WM-basic} accounts for more than 4\,000\,000
lines in the line-force and blocking calculations.

Line blanketing -- i.e. redistribution of flux due to line opacity and
the feedback of this opacity on the temperature structure as a result
of the backwarming effect -- is self-consistently treated in {\sc
cmfgen} and {\sc WM-basic}, but is only accounted for in an
approximate way in the {\sc CoStar} code.  To describe this in some
detail: in {\sc CoStar}, a Monte-Carlo formalism describes the line
transfer problem of the metals. On the one hand, this allows to
describe the feedback of blocking on the atmospheric structure using
wavelength averaged blocking factors. On the other hand, this neglects
photon conversion. This `branching of photons' is relevant for
redistributing flux, as it tends to `soften' the (EUV) radiation
field.

   \begin{figure*}[!p]
   \centering
   \includegraphics[width=15cm]{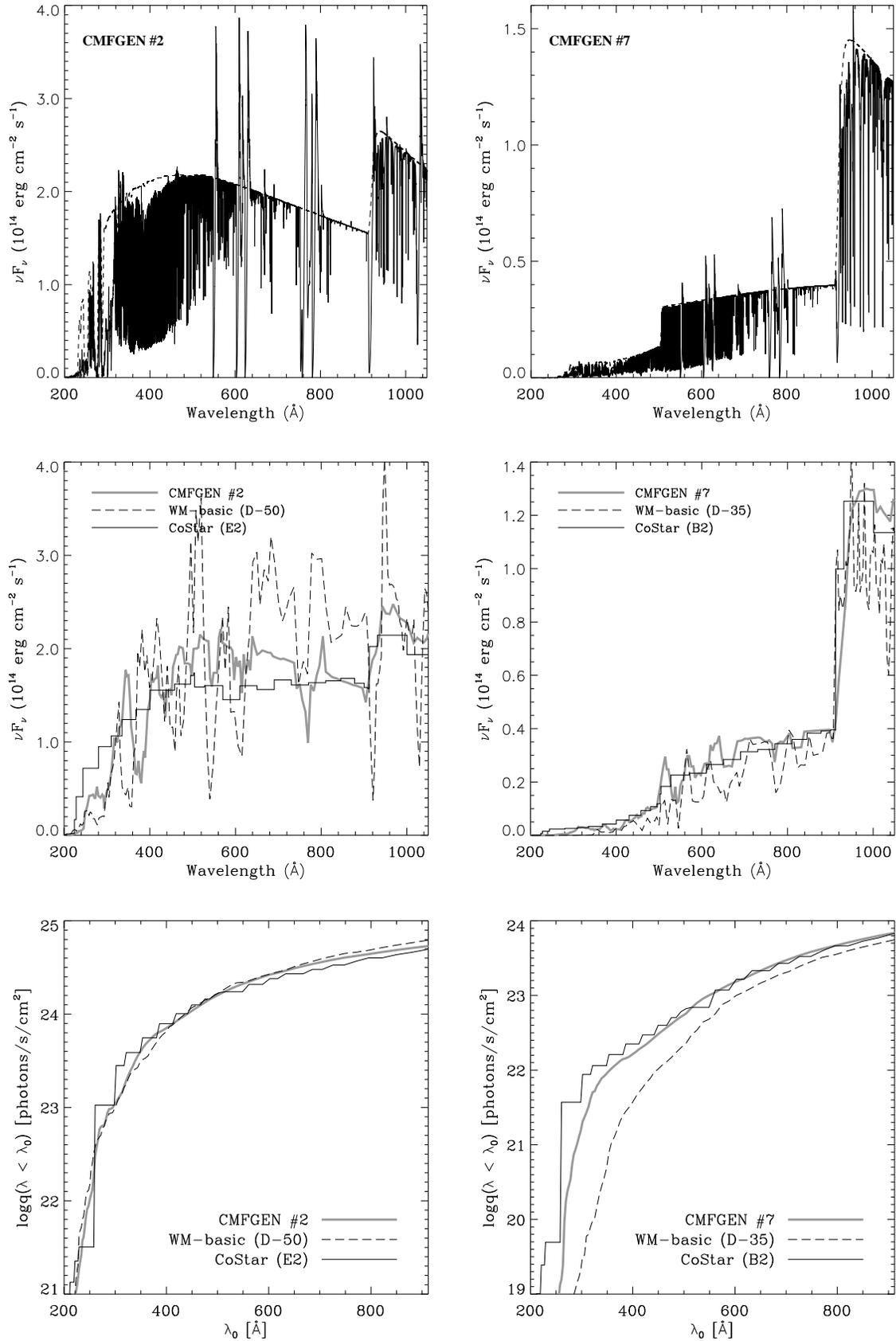}
      \caption{({\it Top panel}) Emergent SED for models \#2 (\Teff\ =
        $48\,670$~K) and \#7 (\Teff\ = $35\,900$~K).  The continuum is
        plotted by a dashed line.  ({\it Middle panel}) Comparison of
        the emergent SED for models \#2 (left panel) and \#7 (right
        panel) with {\sc CoStar} and {\sc WM-basic} dwarf models at
        similar \Teff.  ({\it Bottom panel}) The number of photons in
        cm$^{-2}$\,s$^{-1}$ below wavelength $\lambda_0$ calculated
        from the models considered in the middle panel is plotted as a
        function of this wavelength $\lambda_0$.}
         \label{fig:seds:comparison}
   \end{figure*}

The predicted EUV fluxes (especially in the \HeI~ continuum) are
sensitive to line blanketing. In this respect, the {\sc cmfgen} and
{\sc WM-basic} models include a more consistent treatment and may in
principle be expected to yield more reliable results. For the ionizing
flux, the prescription of the wind structure is relatively unimportant
for main sequence stars, as these have modest winds. The treatment of
blanketing and the specification of the photospheric density structure
affect the strengths of lines, and therefore may have an effect on
\HeI~ and \HeII. As these lines are used to define spectral type, this
may cause a modest uncertainty (up to one sub-type) in the spectral
types assigned to the models. In this respect the {\sc WM-basic}
models are expected to behave best.

\section{The spectral energy distribution}
\label{sect:seds:description}

\subsection{Comparison of predicted EUV fluxes} 

Figure~\ref{fig:seds:comparison} shows a comparison between the
spectral energy distribution of comparable models computed with the
three codes discussed above. The top left panel shows the emergent
ionizing spectrum for our CMFGEN model for \Teff=$48\,670$~K (model
\#2).  In the middle left panel, this {\sc cmfgen} model is shown
together with {\sc WM-basic} model D-50 (\Teff\ = $50\,000$~K,
\citeauthor{pauldrach01} \citeyear{pauldrach01}) and {\sc CoStar}
model C2 (\Teff\ = $48\,529$~K, \citeauthor{schaerer97}
\citeyear{schaerer97}).  The bottom left panel shows partially
integrated $q$ values of these three models. The spectra shown in the
middel left panel are binned in wavelength to better bring out the
overall flux behaviour, which is why they appear smooth. However, a
forest of metal lines does contribute to all of them. The Lyman jump
is modest due to the high degree of ionization. From 912 \AA\ down to
the He\,{\sc i} jump at 504 \AA\ (which is essentially absent),
the most prominent features are strong resonance lines of
     N\,{\sc iv}\,$\lambda 765$, 
     O\,{\sc iv}\,$\lambda 789$,\,$\lambda 609$,\,$\lambda 554$ and
     O\,{\sc v}\,$\lambda 629$.  
In this interval, the {\sc WM-basic} model shows an on average about 
15\% higher flux level compared to our {\sc cmfgen} calculation. This
can be explained by the temperature difference of the models.
Taking \Teff\ effects into account, the {\sc CoStar} model produces a 
somewhat softer spectrum.

In the He\,{\sc i} continuum the spectrum is dominated by a dense
forest of mostly Fe~{\sc iv-vi} lines. Here the {\sc cmfgen} and {\sc
WM-basic} model show a roughly similar behaviour, both in terms of
continuum flux and amount of line blocking. The {\sc CoStar} model now
shows a harder spectrum. The explanation for the {\sc CoStar}
predictions likely involves the treatment of the blocking factors,
which result from a Monte-Carlo simulation in which the effects of
photon conversion (or `branching') are neglected. This forces all
radiative excitation/deexcitation processes to be scatterings, which
prevents flux to be distributed towards longer wavelengths and
therefore keeps the radiation field `hard'. This would also explain
the relatively soft flux in the {\sc CoStar} spectrum at $\lambda
\gtrsim 500$ \AA. Other effects that may play a (more
modest) r\^{o}le are the temperature gradient in the \HeI~
continuum. A steeper gradient causes a higher flux at relatively short
wavelengths (where the continuum is formed at relatively high
Rosseland optical depth), while at relatively long wavelengths the
flux will be lower. Also the neglect of line-broadening in {\sc
CoStar} models may have an effect \citep[see][]{schaerer97}.
 Note that the exact location of the line forest differs between the
{\sc cmfgen} and {\sc WM-basic} models. (To see this properly one
first has to correct the continuum levels for differences in \Teff).
This is most likely connected to differences in the predicted
ionization of iron. The relevant iron ions each show a preferred
region in which their line forest is concentrated. It holds very
roughly that this is at $510 - 440$ \AA\ for Fe~{\sc iv}; $460 - 320$
\AA\ for Fe~{\sc v}; $340 - 260$ \AA\ for Fe~{\sc vi}, and at $270 -
160 $ for Fe~{\sc vii}.  This suggests that the dominant photospheric
ionization of the {\sc WM-basic} model is Fe~{\sc vi}, while Fe~{\sc
v} is most important in {\sc cmfgen} (see the pronounced difference at
$\sim$ 400 \AA).

Though not visible in the figure,
the \HeII~ continuum fluxes of the {\sc cmfgen} and {\sc WM-basic} 
models agree reasonably well (which may be slightly surprising in view 
of the difference in \Teff). Again the {\sc CoStar} model
produces a harder spectrum. The same reasons as discussed above may
apply. However, it may also be connected to the use of the Sobolev 
approximation and/or the velocity structure in the wind.
The latter may cause a depopulation of the \HeII\ ground state
as a result of the velocity gradient in the wind, in combination
with the temperature in the continuum forming layers at the 
wavelength of the \HeII~ resonance lines \citep{gabler89}.

The right panels show a comparison between our {\sc cmfgen} model for
\Teff\ = $35\,900$~K (model \#7), {\sc WM-basic} model D-35 having
\Teff\ = $35\,000$~K, and {\sc CoStar} model B2 for which the
temperature is $36\,300$~K.
Taking into account the differences in effective temperature, the EUV
flux distributions compare well though again the Lyman flux at
$\lambda \gtrsim 700$ \AA\ of the {\sc CoStar} model is somewhat
softer. Below $\sim$ 700 \AA\ -- i.e. again in the regime of severe
flux blocking -- the opposite behaviour is seen: the {\sc CoStar}
spectrum is harder. The likely reasons for this behaviour have been
discussed above. A detailed inspection of the energy distributions
shows that the {\sc WM-basic} models produce a clear drop-off in flux
at the \ion{N}{iii} and \ion{C}{iii} ionization edges near 260
\AA. The presence of this sizable bound-free jump is somewhat
surprising in view of the high \Teff\ of the model, and is not seen in
{\sc cmfgen} and {\sc CoStar} models.


The fluxes in the \HeII\ continua of the three models differ by orders 
of magnitude. Possible reasons for these discrepancies have been
discussed above.

We conclude that down to $\sim$ 500 \AA\ the EUV flux distributions
compare reasonably well, though the flux levels do differ.  However,
at shorter wavelengths large (order of magnitude) discrepancies may
occur. For the {\sc CoStar} predictions, in comparison to the other
two codes, the major part of these differences can be traced back to
the treatment of blanketing.  The models also show large differences
for the \HeII~ continuum fluxes. This is a relevant problem with
respect to the internal consistency of different code
predictions. However, in O-type stars the dominant flux component at
$\lambda < 228$ \AA\ tends to be that of non-thermal soft X-ray
emission arising from shocks in the stellar wind
\citep[e.g.][]{macfarlane94,feldmeier98}.  As no self-consistent
description of this process is presently available, this seems a more
acute problem.

\subsection{Comparison of ionizing fluxes}

   \begin{figure}[!t]
   \centering
   \includegraphics[width=8.5cm]{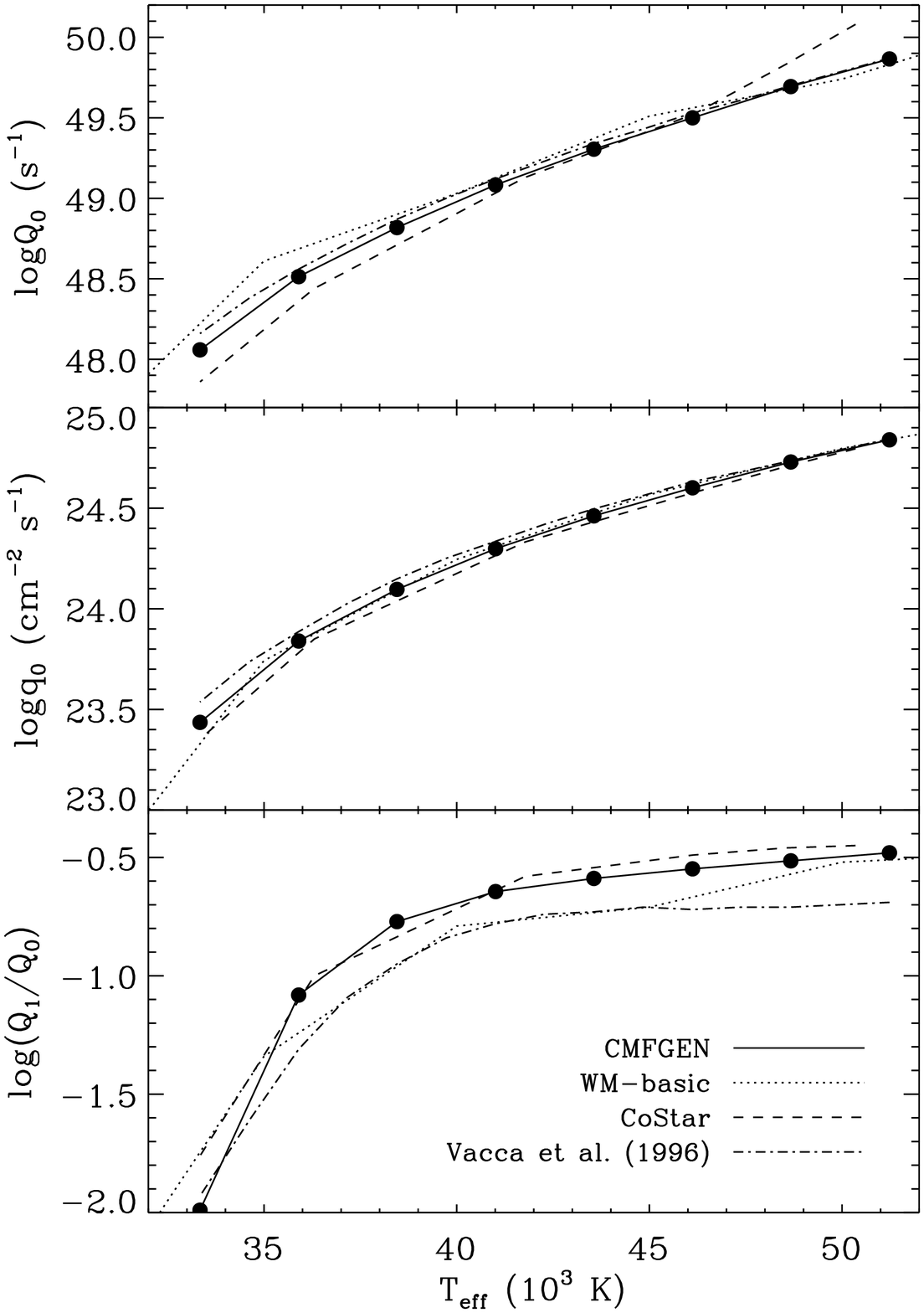}
      \caption{Variation of the Lyman continuum ionizing photons $Q_0$
        ({\it top panel}), $q_0$ ({\it middle panel}), and the
        hardness ratio $Q_1$/$Q_0$ ({\it bottom panel}) with effective
        temperature. The results from our {\sc cmfgen} models (solid
        line) are compared to those from the {\sc WM-basic} models
        (dotted line), {\sc CoStar} models (dashed line) and
        \cite{vacca96} (dash dotted). We use {\sc WM-basic} models
        D-30, D-35, D-40, D-45, D-50 and D-55 \citep{pauldrach01}, and
        {\sc CoStar} models A2, B2, C2, D2, E2 and F2
        \citep{schaerer97}.}
         \label{fig:seds:q}
   \end{figure}

   \begin{figure*}[!t]
   \centering
   \includegraphics[width=15cm]{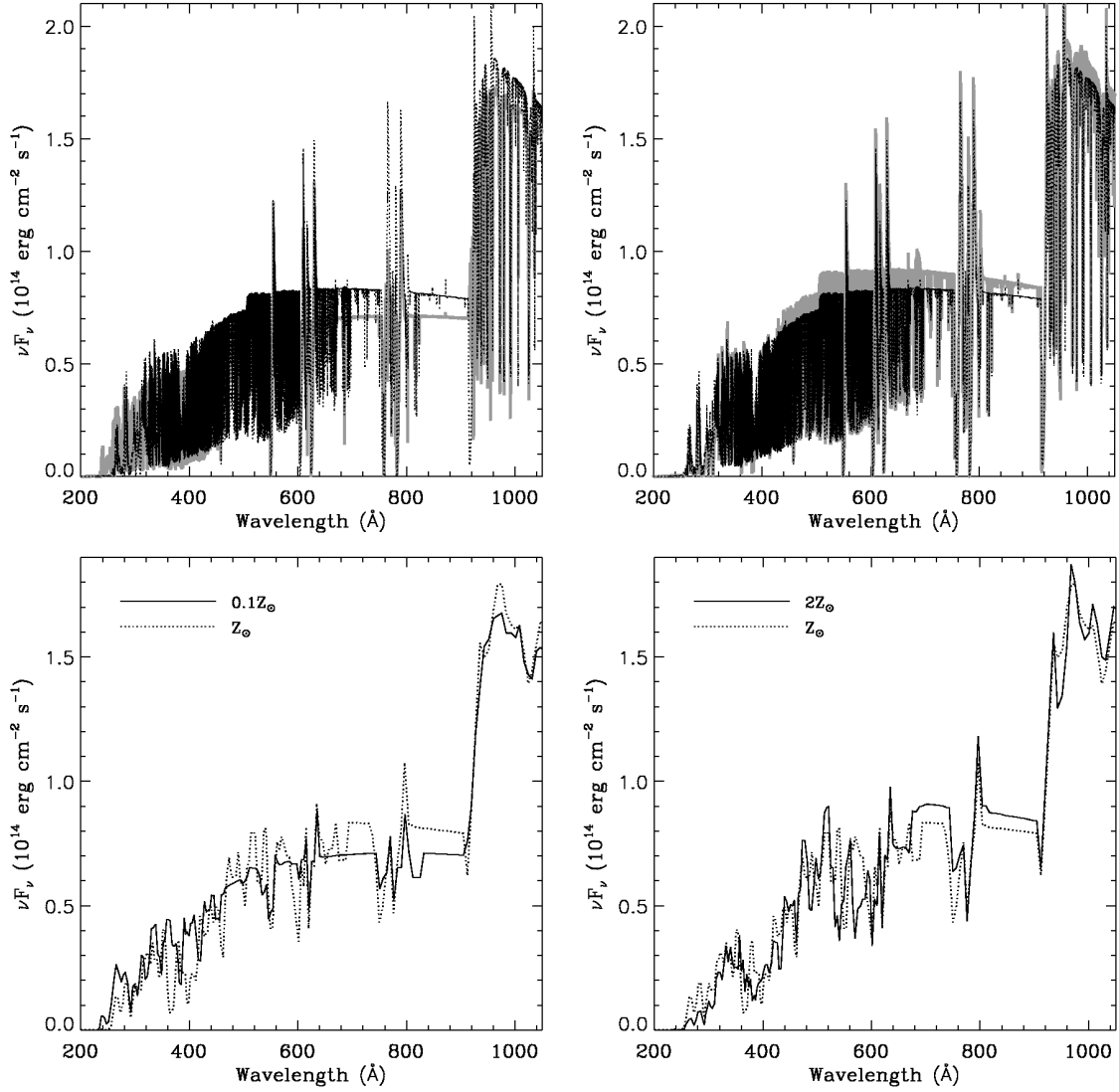}
      \caption{({\it Top panels}) Comparison of the EUV spectrum of
        solar abundance model \#5 (dotted line) with equivalent models
        with $Z=1/10\,Z_{\sun}$ (left panel) and $Z=2\,Z_{\sun}$
        (right panel), plotted in grey. ({\it Bottom panels}) Binned
        version of the spectra plotted in the top panel.}
         \label{fig:seds:sp:z}
   \end{figure*}

Figure~\ref{fig:seds:q} shows the total number of Lyman continuum
photons emitted per second, $Q_{0}$, and the total number per second
per cm$^{2}$, $q_{0} = Q_{0}/4 \pi R_{\ast}^{2}$, for both our O star
main sequence models and the comparison models discussed above. Also
given is the ratio between Lyman and He\,{\sc i} continuum photons,
$Q_{0}/Q_{1}$ (bottom panel). We added predictions from Kurucz models,
presented by Vacca et al. (1996); however, we will not include these
in the discussion of the results.

To eliminate differences in the adopted stellar radii, we first focus
on the predictions for $q_{0}$. The results are plotted versus \Teff,
allowing us to assess the differences in $q_{0}$ as a result of small
differences in adopted temperatures. All predictions are very similar,
only showing discrepancies up to 0.1 dex. Why is this so?

The principal answer is that though line blanketing effects are strong
at wavelengths shorter than $\sim$ 500 (700) \AA\ in the case of model
\#2 (\#7), the flux is mainly redistributed {\em within} the Lyman
continuum. Only limited energy `leaks' out at $\lambda > 912$ \AA.
This may be illustrated using a simple estimate. The line forest
removes $\sim$ 40\% of the flux in both model examples. These photons
are back-scattered and re-thermalized. Assuming the thermal emission
may be characterized by the emerging continuum flux distribution, one
expects 60\% (20\%) of this energy to be emitted still within the
Lyman continuum. Therefore, only 16 \% (32\%) leaks out into the
Balmer continuum and beyond. In terms of number of ionizing photons,
the `damage' is even less as redistribution of energy towards longer
wavelengths within the Lyman continuum requires an increase in the
number of photons.
This explains why the number of ionizing photons only differ up to
$\sim$ 0.05 and 0.1 dex for models \#2 and \#7 respectively. The
neglect of branching in {\sc CoStar} models is separate from the
above discussion (it concerns forward-scattered photons). However,
note that the iron forest is dominated by transitions from meta-stable
levels. The level structure of the relevant iron ions is such that
branching typically produces two photons, one at $\lambda < 912$~\AA\
and one at $\lambda > 912$~\AA. Therefore, also this effect preserves
Lyman continuum photons, though it does affect the hardness ratio
$Q_{1}/Q_{0}$. The former is illustrated in the bottom panels of
Fig.~\ref{fig:seds:comparison}, where we show partially integrated
{\it q} values.

%
%

The hardness ratio $Q_{0}/Q_{1}$, also independent of stellar radius,
yields agreement within $\sim 0.2$ dex. The {\sc CoStar} models having
\Teff\ $>$ $40\,000$~K show a harder spectrum than do the {\sc cmfgen}
and {\sc WM-basic} models -- as discussed above.  This is in agreement
with the flux differences in the Lyman and \HeI~ continua discussed
for model \#2.

The apparent reasonable agreement between number of ionizing photons
is somewhat misleading. Nebular lines from ions such as \ion{Ne}{iii}
or S~{iv} require radiation from below 500 \AA. In this spectral
regime, as we have seen above, the ionizing fluxes may differ strongly
implying large differences in emission line strength.


\subsection{EUV fluxes and $Q_{0}$ and $Q_{1}$ values as a function of
               metal content}
\label{sect:seds:z}  

In Table~\ref{table:seds:q:z} we show the predicted number of ionizing
photons $Q_{0}$ and $Q_{1}$ for models \#2 and \#5 as a function of
metal abundance. The results show that a decrease in $Z$ of a factor
of twenty only alters the number of Lyman continuum photons by 0.03
dex. Over the same metallicity range the number of He\,{\sc i}
continuum photons increases by up to 0.08 dex.

\input{./table5.tex}

\begin{figure}[!t]
   \centering
   \includegraphics[width=8.5cm]{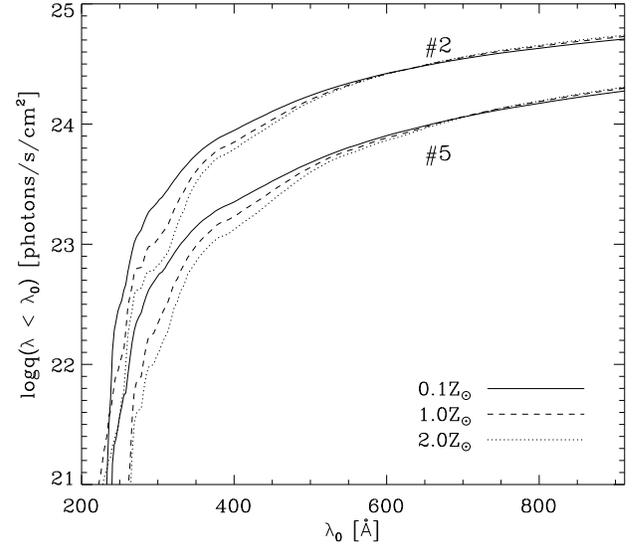}
      \caption{The effect of the metal content on the number of
        photons in cm$^{-2}$\,s$^{-1}$ below a certain wavelength
        $\lambda_0$ is plotted as a function of this wavelength
        $\lambda_0$. The effect is shown for models \#2 and \#5.}
         \label{fig:seds:q:z}
   \end{figure}

The modest dependence on metal content is again the result of the way
line blanketing redistributes the flux (see above). In the top left
panel of Fig.~\ref{fig:seds:sp:z} we show a comparison of the EUV
spectrum of solar abundance model \#5 with that of a model with
identical basic parameters, only with $Z = 1/10 Z_{\odot}$; in the top
right panel the solar and twice solar model are compared. Relative to
the 1/10 $Z_{\odot}$ model, the solar metallicity prediction clearly
shows an increased flux level in the region from the Lyman jump down
to $\sim$ 600 \AA, which is compensated by a decreased flux below 600
\AA.  This is best seen in the bottom left panel, where we show binned
versions of the spectra. The tendency to redistribute flux within the
Lyman continuum is also illustrated in Fig.~\ref{fig:seds:q:z}, where
we show partially integrated $q$ values.

The conservation of ionizing photons works less well for $Q_{1}$, as
expected. Here a decrease in metallicity of a factor twenty results in
up to 20\% more ionizing photons. This implies that those nebular
species that depend on a specific EUV flux range below $\sim$ 500 \AA\
may show a strong dependence on metallicity. We will discuss this in
Sect.~\ref{sect:seds:hii}.

The results for the number of hydrogen ionizing photons are
similar to those obtained by \cite{smith02} using unified {\sc
WM-basic} models, by \cite{lanz2003} using plane parallel TLUSTY
models and by \cite{kudritzki2002}, who used {\sc WM-basic} models of
very massive O stars at $Z=10^{-4}\,Z_{\sun}$ to $Z=Z_{\sun}$. The
results of the two former studies show that the convervation of $Q_0$
for dwarf stars breaks down at lower effective temperatures (at
$32\,000$ to $37\,500$~K depending on model assumptions). Compared to
our models the number of \HeI\ continuum photons in the models of
\citeauthor{smith02} do not show a similar behaviour. The models of
\citeauthor{kudritzki2002} and \citeauthor{lanz2003} do show a
conservation of the number of photons, where the latter models again
show that this convervation breaks down for lower effective
temperatures ($\lesssim 45\,000$~K).

\section{Spectral classification}
\label{sect:seds:sptype}

\subsection{Optical classification}   
\label{sect:seds:cal_opt}

The optical spectral classification of O stars relies on a {\em
quantitative} criterion based on the ratio of the equivalent widths of
\HeI~$\lambda$4471 and \HeII~$\lambda$4542
\citep{conti71,conti77,mathys88}. In O9 stars, the \HeI~$\lambda$4471
line is stronger; the ratio changes gradually so that the
\HeII~$\lambda$4542 line begins to dominate at type O6 and earlier.
According to this criterion, the value of
log$[W_\lambda(4471)/W_\lambda(4542)]$ unambiguously determines the
spectral type of the star.

\begin{figure*}[!t]
   \centering
   \includegraphics[width=1.7\columnwidth]{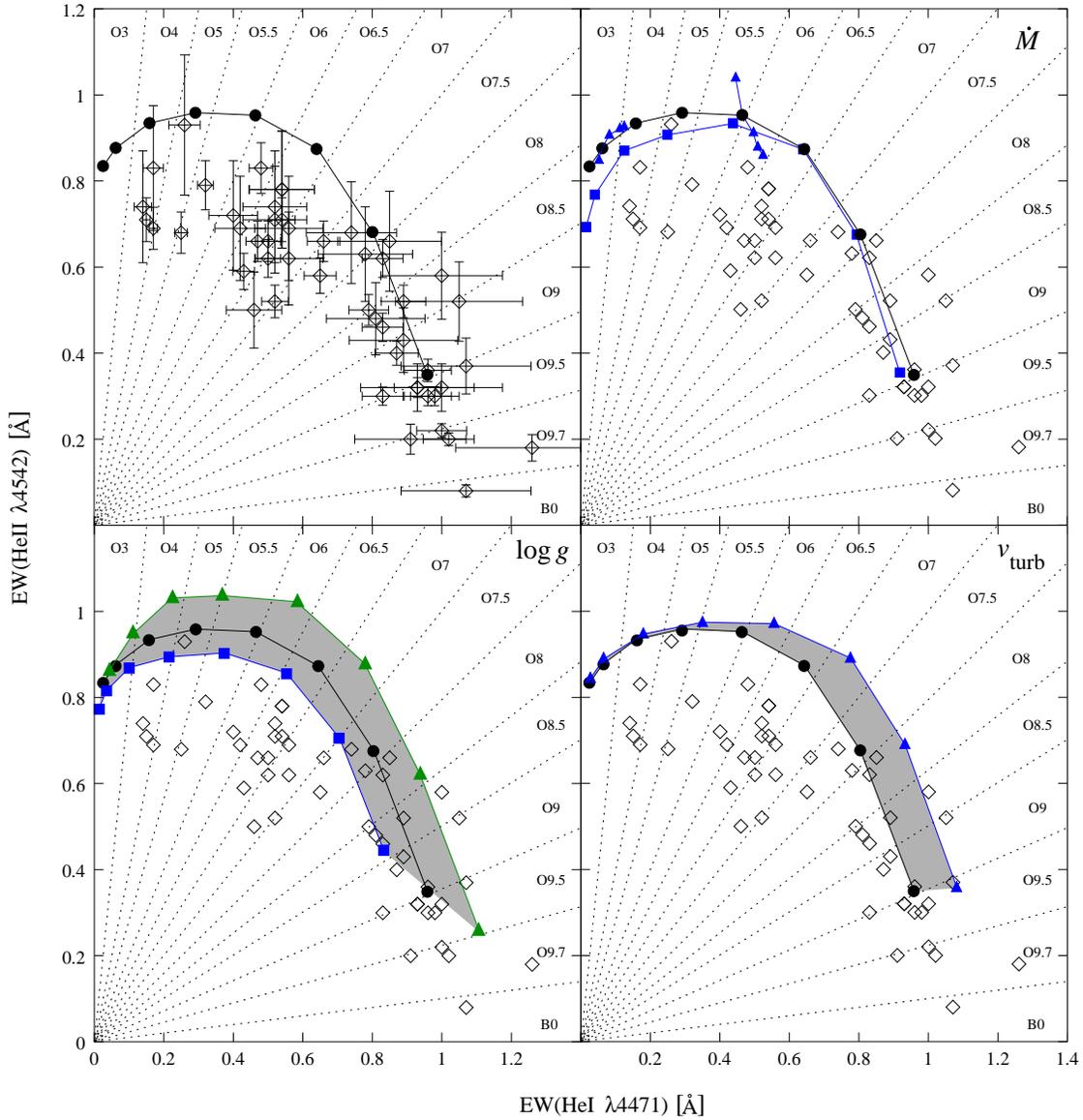}
      \caption{Top left panel compares the equivalent width of
      \HeI~$\lambda$4471 and \HeII~$\lambda$4542 measured in the
      model grid with observed values in main sequence (dwarf) stars
      (open diamonds). Bottom left panel shows the effects of an
      increase (triangles) and a decrease (squares) of 0.3 dex in
      $\log g$. In the top right panel a grid with doubled mass-loss
      (squares) is compared to the model grid (circles). Also shown
      are the equivalent widths of models \#2 and \#5 with $Z=2, 1/2,
      1/5$ and 1/10\,$Z_{\sun}$ (triangles). The decrease in
      metallicity results in an increase of the equivalent width of
      \HeI\ 4471 \AA. The bottom right panel compares a grid with
      $v_{\rm turb}= 5$~\kms\ (circles) to a grid with $v_{\rm turb}=
      20$~\kms\ (triangles). The dotted lines indicate the parameter
      space delimited by every subclass according to the empirical
      classification by \cite{mathys88}.}
      \label{fig:seds:ew_opt}
   \end{figure*}

This observational criterion can be used to assign spectral types to
our models. Fig.~\ref{fig:seds:ew_opt} compares the equivalent widths
of the \HeI\ and \HeII\ classification lines observed in dwarf stars
with the predictions from our models. The slopes of the dotted lines
correspond to the ratios of the equivalent widths delimiting the O
star subclasses according to the calibration of \cite{mathys88}. We
investigate the assigned spectral classification for the effects of
varying respectively $\log g$, metallicity and microturbulent
velocity. Before discussing these dependencies in detail, we will
first elaborate on how well our initial model grid for dwarf stars
fits these observations.

The observational data for the dwarf stars are taken from
\cite{conti71}, \cite{conti73}, \cite{conti77}, and
\cite{mathys88,mathys89}. The luminosity class was verified using the
spectral classification of \cite{walborn1971, walborn1972,
walborn1976, walborn1973, walborn1982}. Stars that were not classified
as dwarfs by Walborn were discarded. Spectroscopic binaries were
identified using the catalogue of \cite{pedoussaut1996} and stars with
variable radial velocity using \cite{mathys88,mathys89}. These were
also removed. Finally, variable stars were deselected using the
catalogue of \cite{kholopov1998}. In the top left panel of
Fig.~\ref{fig:seds:ew_opt} we show the selected observed data (using
open diamond symbols), including the error bars. The observations
cover a strip in the \HeI~$\lambda$4471 versus \HeII~$\lambda$4542
equivalent width diagram that is broader than can be accounted for by
uncertainties in individual measurements. The width of this strip
therefore also reflects differences in stellar properties such as
gravity, mass loss, metallicity and/or photospheric properties such as
microturbulent velocity.

\subsubsection{Mass loss}
Also plotted in the top left panel are the models listed in
Table~\ref{table:seds:grid}. The predictions reproduce the observed
range of \HeI~$\lambda$4471 equivalent width values well. The
equivalent width of \HeII~$\lambda$4542 seems to be overestimated by
approximately 10 to 20\% for early- and mid-O spectral types.  A
possible explanation for this overprediction may be an adopted mass
loss that is systematically too low. A higher mass loss would reduce
the equivalent width, as wind emission fills in more of the
photospheric absorption profile \citep{lanz96}. This effect is
especially important for the earliest spectral types, as these have
the strongest winds. The top right panel of Fig.~\ref{fig:seds:ew_opt}
shows the effect of an increase in $\dot{M}$. Here a grid with doubled
mass loss rate, indicated by squares, is compared to the standard grid
(circles). A change of this magnitude is characteristic of the typical
error in both mass-loss determinations \citep{puls1996} and mass-loss
predictions \citep{vink00}. The line strength of \HeII~$\lambda$4542
is decreased by up to 10 to 20\% in the early type models, as a result
of wind emission. A modest decrease in the \HeI~$\lambda$4471 strength
is also observed.

\subsubsection{Gravity}
The width of the observed strip of equivalent width ratios is partly
explained by gravity effects. \cite{kudritzki83} already showed that
the ratio $W_\lambda(4471)/W_\lambda(4542)$ depends on luminosity
class. Here we show that gravity variations {\em within a luminosity
class} also significantly effect this ratio. The main sequence
gravities adopted by Vacca et al. (1996) should be seen as mean values
for the ensemble of stars of given spectral type used for their
calibration. We have calculated a grid of models in which we increased
and decreased the stellar gravity by 0.3 dex. This roughly corresponds
to the typical difference between dwarf and giant stars. The bottom
left panel of Fig.~\ref{fig:seds:ew_opt} shows the predicted strip of
equivalent width ratios spanned by this gravity range. Note that e.g.\
a decrease (increase) of $\log g$ by 0.3 dex implies about one
spectral sub-type earlier (later). A decrease of $\log g$ implies a
decrease in electron density, which causes a decrease in the
recombination rate, effectively increasing the helium ionization. The
opposite is true in case of an increased gravity.

The new temperature scale of O stars (\citealt{martins02}; see also
\citealt{dekoter98}) implies a decrease in \Teff\ of about 10 percent
compared to the calibration of \cite{vacca96} that is used for our
grid. The lower luminosity and mass implied by this new \Teff\
results, however, in only a modest shift of $\sim$~0.06 dex in
gravity. Therefore, this temperature recalibration can only account
for a very minor part of the overprediction of \HeII~$\lambda$4542.

In part the overprediction of \HeII~$\lambda$4542 may also be the
result of a to high stellar mass. Indeed, a systematic difference has
been noted in masses that are based on spectroscopic analysis of
Balmer line profiles an those that result from a comparison of basic
parameters ($L_\ast$ and \Teff) with evolutionary tracks
\citep[e.g.][]{herrero92}. Recent new results \citep{herrero2002}
appear to show that most of this discrepancy can be explained by
improved spectroscopic modeling and the lower \Teff\ values for
O-stars. Still, adopting the spectroscopic masses as presented by
\cite{vacca96} (which are 30 to 70\% less for the earliest spectral
types) yields a decrease in \HeII~$\lambda$4542 equivalent width of
about 10 to 15\% for the earliest spectral types. This should be seen
as an upper limit of a possible mass effect.

\subsubsection{metallicity}
metallicity effects may also explain part of the scatter in equivalent
widths, but are not expected to be responsible for the systematic
shift in the \HeII~$\lambda$4542 line strength.  Variations in the
stellar metallicity are undoubtedly present in Galactic O stars.
Emission-line studies of \HII\ regions, planetary nebulae and
supernova remnants consistently show that the average abundance of
oxygen decreases with Galactocentric distance by $\sim 0.06$ dex per
kpc, with values of $\sim 2.5\,Z_{\sun}$ in the Galactic Center, and
down to $\sim 1/4\,Z_{\sun}$ at Galactocentric distances larger than
15 kpc \cite[see e.g.][]{henry99}. Low abundance systems such as the
Small Magellanic Cloud have metallicities of the order of $\sim
1/8\,Z_{\sun}$.  The top right panel in Fig.~\ref{fig:seds:ew_opt}
shows the variation of the equivalent widths of \HeI~$\lambda$4471 and
\HeII~$\lambda$4542 as a result of changing the atmospheric
metallicity -- and its indirect effect on line blanketing and stellar
wind mass loss -- from 1/10 to 2 $Z_{\sun}$. The models with modified
metal content are indicated by solid triangles.

Two effects related to metal content may influence the line strength
of the \HeI~$\lambda$4471 and \HeII~$\lambda$4542. First, an increased
metallicity implies an increased line blanketing and therefore
stronger diffuse field. This will increase the ionization of
helium. The top right panel of Fig.~\ref{fig:seds:ew_opt} indeed shows
this behaviour for model \#5, i.e.\ a decrease in the \HeI\ and an
increase in the \HeII\ line strength. Second, an increased metallicity
implies a stronger stellar wind, therefore an increased filling in of
line profiles by wind emission \citep[see also
e.g.][]{herreroIAUS212}. This effect is more important for earlier
type stars, which have a larger mass loss. This second effect
dominates in model \#2, resulting in a decreased \HeII~$\lambda$4542
line strength with increased metallicity and vice versa.

We conclude that metallicity effects may change the spectral type by
up to one subtype. This is in agreement with findings of
\cite{pulsIAUS212}, which showed that the inclusion of line blanketing
at SMC metallicities resulted in a much more modest change in \Teff\
relative to that found for Galactic stars.


\subsubsection{Microturbulence}
The effect of microturbulence is expected to explain part of the
scatter in the observed \HeI~$\lambda$4471 and \HeII~$\lambda$4542
equivalent width ratios and is shown in the bottom right panel of
Fig~\ref{fig:seds:ew_opt}. The standard grid assumes a constant
microturbulent velocity of 5~\kms. We calculated a new grid adopting
$v_{\rm turb} = 20$~\kms, producing stronger lines. A comparison shows
that the \HeII\ line is almost insensitive to the introduced variation
in microturbulent velocity. The saturated \HeI\ line however shows a
significant change in strength for models of spectral type O5 and
later. The relative changes are 4\% for the model with the weakest
\HeI\ line and 12\% to 20\% for the models with the strongest \HeI\
lines. An increased microturbulence may shift the assigned spectral
type by up to half a sub-type towards later spectral type. The impact
of microturbulence in the line formation calculations of H and \HeI\
in OB stars, and consequently, in the derived stellar atmospheric
parameters, has been investigated by several authors
\citep[e.g.][]{smith98,mcerlean98,villamariz00}.

Summing up the above discussion, we conclude that all-in-all the
predicted equivalent width of \HeI~$\lambda$4471 and
\HeII~$\lambda$4542 reproduce the observed behaviour reasonably well,
though the \HeII\ line strength is somewhat overpredicted for early-
and mid-O types. We also show that the spectral type of dwarf stars is
not only a function of effective temperature of the star, but also
depends, albeit to a lower degree, on microturbulent velocity, metal
content and on gravity, i.e.\ gravity variations within the dwarf
class. Variations in these parameters explain the observed strip of
equivalent width of \HeI~$\lambda$4471 and \HeII~$\lambda$4542 for
given spectral type.

   \begin{figure}[!t]
   \centering
   \includegraphics[width=8.5cm]{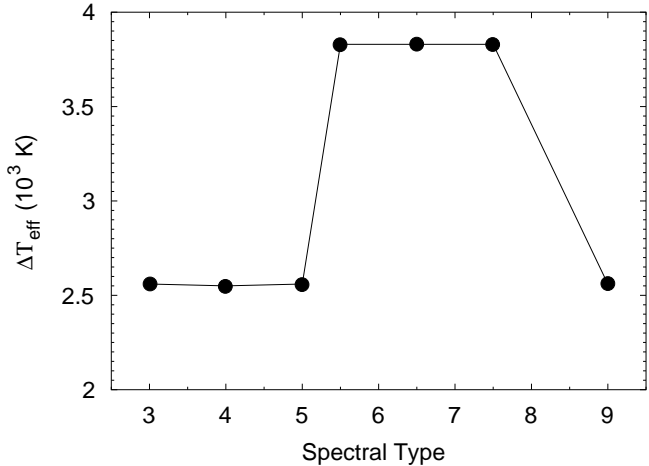}
      \caption{Effective temperature shift with spectral type between
        the \cite{vacca96} calibration and the temperature scale
        derived from the models with $Z=Z_{\sun}$ and $v_{\rm turb}=
        5$~\kms.}
         \label{fig:seds:shift}
   \end{figure}

Finally, when comparing the spectral types of our models to the
original spectral types assigned by \cite{vacca96}, we see that there
is a systematic shift to earlier spectral types for given
temperature. In other words, the \Teff\ scale derived with our models
shows a systematic shift to lower temperatures.
Fig.~\ref{fig:seds:shift} shows the difference for $Z=Z_{\sun}$ and
$v_{\rm turb} = 5$~\kms. The decrease from the \citeauthor{vacca96}
scale to our new scale varies between $\sim 2500$~K and $\sim 3800$~K
and is largest for stars in the spectral range of O5 to O9. This
decrease of \Teff\ due to line blanketing effects has recently been
found in several studies
\citep[e.g.][]{dekoter98,crowther01,martins02}.

\subsection{K-band classification}
\label{sect:seds:cal_k}

K-band (2 \mum) spectroscopy has been used in the past few years to
identify and classify the characteristics of newly-formed O stars that
are found in heavily obscured \HII\ regions
(e.g. \citeauthor{watson97a} \citeyear{watson97a},
\citeauthor{watson97b} \citeyear{watson97b}, \citeauthor{kaper02a}
\citeyear{kaper02a}, \citeauthor{bik03} \citeyear{bik03},
\citeauthor{martin:G29} \citeyear{martin:G29}).  The typical dust
extinction of more than 10 mag at visual wavelengths towards these
\HII\ regions has restrained optical photometry and spectroscopy of
such stars. At near infrared wavelengths, the extinction is much less,
and direct observations of their photospheres are possible.  At longer
wavelengths, observations fail to detect their photosphere directly
because the spectral energy distribution is dominated by emission from
the surrounding dust. Hence, it becomes essential to understand the
behaviour of the photospheric features in the near-IR regime, the only
observational window of such young O stars. In this section we
compare the K-band spectral behaviour of our models with observations.

\cite{hanson96} present an atlas of K-band spectra of a large number
of relatively nearby, well-studied, optically visible massive stars
with the intention of investigating the variation of 2 \mum\ spectral
features with spectral type and luminosity in known OB stars. They
find that in early-O dwarf stars (O3, O4, some O5),
\HeII~$\lambda$2.189 (wavelengths of the near-IR lines are in
micrometers) absorption and \kNIII\ (observed at 2.116 \mum) emission
are the dominant spectral indicators. Mid-O stars (O5, O6, some O7)
show \kCIV~$\lambda$2.078 emission and can begin to show
\HeI~$\lambda$2.113 absorption. Late-O (O8, O9) and early-B stars lack
\HeII, \kNIII\ and \kCIV, showing instead \HeI\ and fairly strong (in
comparison to hotter O stars) \HI~$\lambda$2.1661 and
Brackett~$\gamma$.

Our grid of models can be used as a first test against the features
found in the K-band spectrum of O-type dwarf stars. Many complications
are to be expected in the modeling of these lines since they are
formed in the transition region from the stellar photosphere to the
super-sonic stellar wind \citep[see e.g.][]{Kudritzki2000}. A full
parameter study is highly desirable to understand their diagnostic
value. In this paper, we only discuss the temperature and
microturbulent velocity effects and refer to Lenorzer et al.\ (in
prep.) for a detailed parameter study.

   \begin{figure}[!t]
   \centering
   \includegraphics[width=8.5cm]{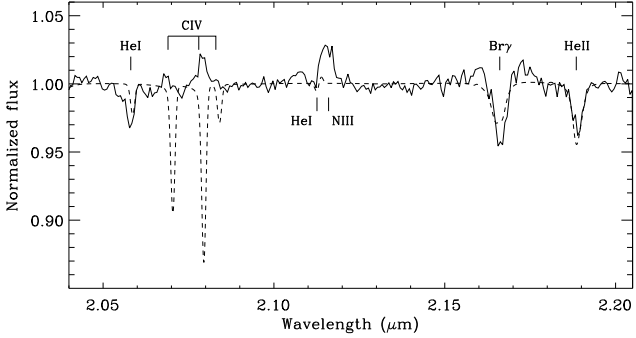}
      \caption{The synthetic K-band spectrum from model \#4 with
        $Z=Z_{\sun}$ and $v_{\rm turb} = 10$~\kms, which corresponds
        to an O5 main sequence star (dashed line) is
        compared to the spectrum of the O5V((f)) star HD\,93204 (solid
        line). The synthetic spectrum is degraded to the resolution of
        the observed spectrum ($\lambda/\Delta\lambda \sim 1500$). The
        lines observed in the K-band are indicated. Note that while
        the \ion{C}{iv} doublet transition is clearly observed in
        emission, our model predicts it to be in absorption. The
        \kNIII\ emission at 2.116 \mum\ is not included in the
        models.}
         \label{fig:seds:kband:sp}
   \end{figure}

Figure ~\ref{fig:seds:kband:sp} shows, as an example, the comparison
between the synthetic K-band spectrum from model \#4 with $Z=Z_{\sun}$
and $v_{\rm turb} = 10$ \kms~ (which corresponds to an O5 main
sequence star) and the spectrum of HD\,93204, an O5V((f)) star
observed by \cite{hanson96}. The synthetic spectrum was degraded to
the resolution of the observed spectrum ($\lambda/\Delta\lambda \sim
1500$). The \HeI\ lines at 2.058 and 2.113 \mum, Br$\gamma$, and the
\HeII\ line at 2.189 \mum\ are clearly visible in the synthetic
spectrum. Qualitatively, the behaviour of these lines closely
resembles that shown in the 2 \mum\ atlas. \HeI~$\lambda$2.058 is
almost undetectable in our three hottest models but starts to show a
strong absorption in the intermediate temperature models to finally
weaken in model \#8. \cite{hanson96} observed that this line may turn
into emission for B stars. Br$\gamma$ and \HeI~$\lambda$2.113
strengthen towards later-type stars, while \HeII~$\lambda$2.189
becomes progressively weaker.

The models show a discrepant behaviour in the case of the metal
lines. The \kNIII\ emission at 2.116~\mum, with transition 8--7, is
simply not included in the models. The \kCIV\ doublet transition
(3pP$^0-$3d$^2$D), observed in emission O4 to O8.5 stars
\citep{hanson96}, is accounted for in our models. It is, however,
predicted in absorption in all models that have $\Teff \gtrsim 41\,
000$~K, i.e.\ spectral type O5 or earlier. The disappearance of these
lines in late O-type stars is well predicted and is the result of the
recombination of \kCIV\ to \kCIII. The origin of the observed emission
lines is more difficult to understand. Recent high spectral resolution
($R\sim 8000$) observations of O-type stars (Bik et al.\ in prep.)
resolve these \kCIV\ lines in a few stars, yielding a FWHM of about
40~\kms. Moreover, the strength of the lines (e.g.\ Br$\gamma$) as
observed by \citeauthor{hanson96} is not found to be much stronger in
supergiants than in dwarf stars. These arguments strongly suggest that
the near-infrared \kCIV\ lines have a predominantly photospheric
origin, therefore their emission cannot be attributed to an enhanced
\kCIV\ abundance and/or wind effects. If the emission was due to a
temperature inversion in the line forming region, this would
similarily affect \HeI~$\lambda$2.058. Indeed, both lines have regions
of formation that largely overlap such that a temperature inversion
would also be visible in the \HeI\ line. Our models, however,
reproduce the \HeI\ line reasonably well. The transition of \kCIV\
observed in the near-infrared spectrum of O-stars involves low atomic
levels that are populated via transitions located in the UV and
far-UV. It is therefore likely that we do not reproduce the behaviour
of these lines due to a mistreatment of this complex spectral
region. This issue is still being investigated but is encouraged by
recent calculations including \kCIII\ which show the \kCIV\ doublet
transition in emission. We therefore suspect that line pumping is the
cause of these emission lines (Lenorzer et al.\ in prep.)

\begin{figure}[!t]
   \centering
   \includegraphics[width=8.5cm]{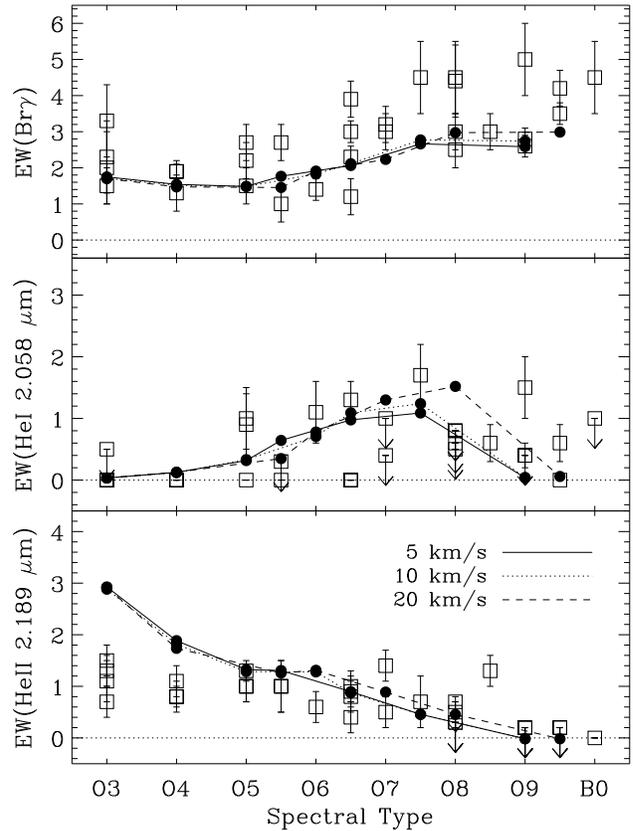}
      \caption{Comparison between observed equivalent widths of
       Br$\gamma$, \HeI~$\lambda$2.058 and \HeII~$\lambda$2.189 in
       main sequence stars (open squares, \citeauthor{hanson96}
       \citeyear{hanson96}) and those measured from our models with
       $v_{\rm turb}$ = 5 (solid line), 10 (dotted line) and 20~\kms
       (dashed line).}
         \label{fig:seds:ew:kband}
   \end{figure}

We carried out a quantitative comparison with observed equivalent
widths using the \HeI~$\lambda$2.058~, Br$\gamma$ and
\HeII~$\lambda$2.189. Fig.~\ref{fig:seds:ew:kband} compares the
predicted equivalent widths of these three lines adopting $v_{\rm
turb}$ = 5, 10 and 20~\kms, with those measured by
\cite{hanson96}. Spectral types of the models are those derived from
the optical lines (see Fig.~\ref{fig:seds:ew_opt}). The microturbulent
velocity has no significant effect on the lines. The models fit
reasonably well the observed equivalent widths, although they
overestimate the strength of \HeII~$\lambda$2.189 for spectral types
O3 and O4.

We can investigate whether the equivalent widths of the 2.058 \mum\
\HeI, Br$\gamma$ and 2.189 \mum\ \HeII\ lines are affected by
variations in the metal content. For model \#2 with $v_{\rm turb} =
10$~\kms\ (which corresponds to an O3 star), a change in metallicity
from 2 to 1/10\,$Z_{\sun}$ produces a decrease in the equivalent width
of the \HeII\ line from 3.3 to 1.8 \AA. Br$\gamma$ shows only
variations at the 30\% level and the \HeI\ line is practically absent
at all metallicities (equivalent width $< 0.1$~\AA). In the case of
model \#5 with $v_{\rm turb} = 10$~\kms\ (an O6 star), the equivalent
width of the \HeI\ line remains practically unchanged; the Br$\gamma$
and \HeII\ line show variations of the order of 20--30\%.

Summing up, variations in microturbulent velocity have no significant
impact on the equivalent width of the Br$\gamma$, \HeI\ and \HeII\
lines, while variations in metal content have an appreciable influence
on the equivalent width of the \HeII\ line in the case of very hot
stars.

   \begin{figure*}[!t]
   \centering
   \includegraphics[width=12cm]{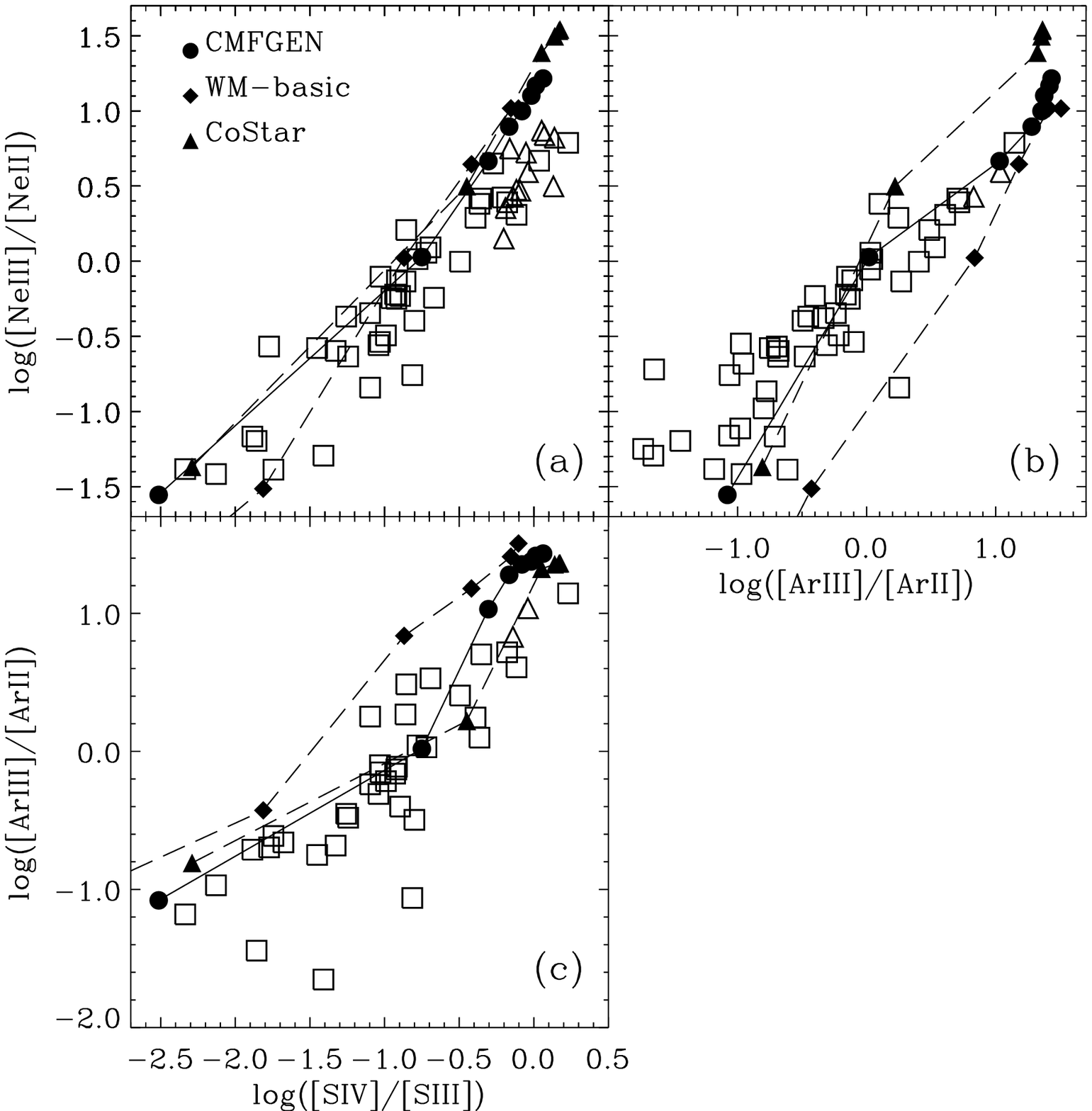}
      \caption{Comparison of the infrared fine-structure line ratios
       \arratio~9.0/7.0, \sratio~10.5/18.7 and \neratio~15.5/12.8
       \mum\ observed in Galactic (open squares) and Magellanic Cloud
       (open triangles) \HII\ regions with predictions from
       photoionization models at {\it solar metallicity} based on
       three different SEDS: {\sc cmfgen} (solid circles), {\sc
       WM-basic} (solid diamonds) and {\sc CoStar} (solid triangles).
       A unique ionization parameter, $U$, is adopted for all the
       models. We stress that unless $U$ is constrained, these line
       ratios cannot be used to derive an absolute value for \Teff.}
         \label{fig:seds:cloudy}
   \end{figure*}

\section{The ionizing structure of single star \HII\ regions}
\label{sect:seds:hii}

Observations of \HII\ regions combined with detailed photoionization
models can be used to test the EUV spectrum of O stars. In this
respect, the Infrared Space Observatory (ISO, 
\citeauthor{kessler96} \citeyear{kessler96}) provided an unique
opportunity by measuring the infrared (2.3--196~\mum) spectra of a
large sample of \HII\ regions
\citep{giveon02,peeters:catalogue,vermeij:data}, giving access to four
elements (N, Ne, S and Ar) in two different ionization stages. Ratios
of fine-structure lines such as \NIII/\NII\ 57/122~\mum, \NeIII/\NeII\
15.5/12.8~\mum, \SIV/\SIII\ 10.5/18.7~\mum\ and \ArIII/\ArII\
9.0/7.0~\mum\ probe the ionizing stellar spectrum between 27.6~eV (450
\AA) and 41~eV (303 \AA) and can be used as indicators of the realism
of stellar models at these energies.

Basically, the ratio of two successive stages of ionization $X^{\rm
+i}$ and $X^{\rm +i+1}$ of a given element $X$ indicates the state of
ionization of the nebula, which depends principally
on the shape of the SED, more
specifically on the number of photons able to ionize $X^{\rm +i}$
compared to that of Lyman continuum photons \citep{vilchez88}, and on
the ionization parameter $U$. 
In the case of an ionized sphere of constant gas
density, $n$, and filling factor, $\epsilon$, $U$ is defined by  
$U=Q_{0}/(4\pi R_{\rm s}^2 nc) \propto (Q_{0}n\epsilon^2)^{1/3}$ 
\citep[see e.g.][]{stasinska97}, where $R_{\rm s}$ is the radius of the
Str\"{o}mgren sphere and $c$ is the speed of light. Any combination of
$Q_{0}$, $n$ and $\epsilon$ that keeps the product $Q_{0}n\epsilon^2$
constant results in a similar ionization structure.

In order to investigate the dependence of the fine-structure line
ratios on the SED, nebular models have been computed with the
photoionization code CLOUDY version 96.00-beta\,4\footnote{see
http://thunder.pa.uky.edu/cloudy/} using MICE, the IDL interface for
CLOUDY created by H. Spoon\footnote{see
http://www.astro.rug.nl/$\sim$spoon/mice.html}. We computed nebular
models for static, spherically symmetric and homogeneous gas
distributions with one ionizing star in the center. An inner cavity
with a radius equal to 10$^{17}$ cm is set, while the outer radius of
the \HII\ region is defined by the position where the electron
temperature reaches $1\,000$~K. The grid of stellar models based on the
{\sc cmfgen} code and described in Sect.~\ref{sect:seds:models} is
taken to describe the incident SED. We take models with $v_{\rm turb}
= 10$~\kms.  The stellar spectra were rebinned to 2\,000 points, which
is the limit of points for the input SED accepted by CLOUDY.  The
nebular metallicity was set equal to that of the stellar model used to
calculate the SED.  Finally, with the aim to explore the dependence of
the fine-structure line ratios on the SED alone, we decided to fix the
ionization parameter $U$, so that the variations of the line ratios
are only due to variations of the SED appearance. Hence, we fix $Q_0$
to the typical value of $10^{48}$~photons~s$^{-1}$, the density to
$n=10^3$~\cm3, and the filling factor to $\epsilon=1$.  We stress here
that unless $U$ is constrained, these line ratios cannot be used to
derive an absolute value for \Teff\ \citep{stasinska97}.

We computed several sets of nebular models. First, a set at solar
metallicity where the dependence on effective temperature is analyzed,
and second, a set where the effective temperature is fixed and the
metal content is modified. This last grid allows us to investigate the
dependence of the fine-structure line ratios on metallicity.

Figure~\ref{fig:seds:cloudy} presents the correlations observed
between the fine-structure line ratios \ArIII/\ArII\ 9.0/7.0~\mum,
\SIV/\SIII\ 10.5/18.7~\mum\ and \NeIII/\NeII\ 15.5/12.8~\mum\ for the
sample of Galactic \citep{giveon02,martin:paperii} and Magellanic
Cloud \HII\ regions \citep{vermeij:data} observed by ISO.  The
ionization potentials of \ion{Ar}{ii}, \ion{S}{iii} and \ion{Ne}{ii}
are, respectively, 27.6 eV (450 \AA), 34.8 eV (357 \AA) and 41.0~eV
(303 \AA). We do not include the ratio \NIII/\NII\ 57/122~\mum\ in
this figure; the ionization potential of \ion{N}{ii} (29.6 eV) is
close to that of \ion{Ar}{ii}, and hence, this ratio roughly probes
the same range of energies as \ArIII/\ArII.  The data span a range in
ionization of more than two orders of magnitude. Interestingly, the
Magellanic Cloud \HII\ regions nicely overlap with the trend observed
for the Galactic objects at the high ionization end
\citep[see][]{martin:metal}.
 
\begin{figure}[!t]
   \centering
   \includegraphics[width=8.5cm]{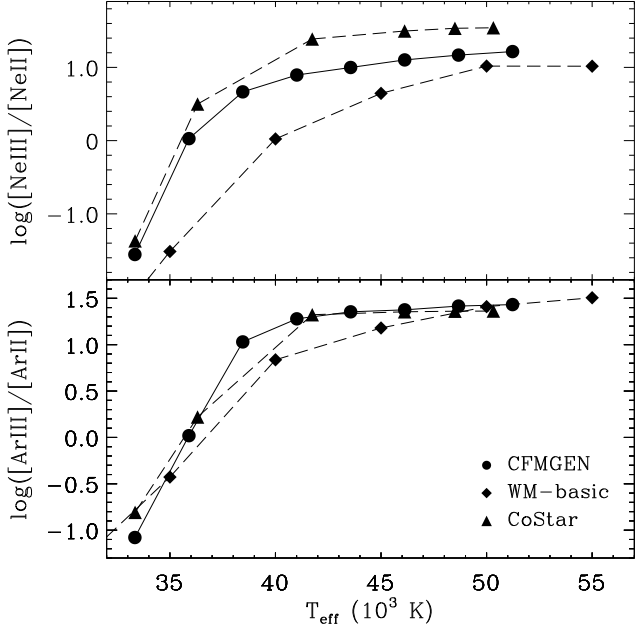}
      \caption{Predicted \ArIII/\ArII\ 9.0/7.0 \mum\ and \NeIII/\NeII\
        15.5/12.8 \mum\ fine-structure line ratios as a function of
        \Teff\ for the three sets of models at solar metallicity
        considered in the text. A unique ionization parameter, $U$,
        is adopted for all the models. We stress that unless $U$ is
        constrained, these line ratios cannot be used to derive an
        absolute value for \Teff.}
      \label{fig:seds:cloudy:teff}
   \end{figure}

The diagrams in Fig.~\ref{fig:seds:cloudy} probe the shape or hardness
of the stellar spectrum between 303 and 357~\AA\ (panel a), 303 and
450~\AA\ (panel b), and 357 and 450~\AA\ (panel c). In this diagram
the results for the set of nebular models at solar metallicity are
confronted to the observations. As a comparison, we also show the
results of grids of models computed using the {\sc WM-basic} and {\sc
CoStar} stellar models. We use {\sc WM-basic} models D-30, D-35, D-40,
D-45, D-50 and D-55 \citep{pauldrach01}, and {\sc CoStar} models A2,
B2, C2, D2, E2 and F2 \citep{schaerer97}. For all model sets, the
predicted nebular excitation increases with increasing effective
temperature of the ionizing star. However, differences among the
models, and between the models and the observations, are clearly
present. The largest differences are seen in panels (b) and (c), for
which the range of energies tested is larger.  Several conclusions can
be gleaned from the inspection of Fig.~\ref{fig:seds:cloudy}.  (1) For
the same nebular conditions, i.e. same ionization parameter $U$, the
nebular models based on the {\sc CoStar} SEDs with \Teff\ $>
36\,000$~K largely overpredict \neratio\ for a given \sratio\ or
\arratio\ when compared to the observations and the other SEDs. This
can be understood in terms of an excess of photons below 303 \AA. This
effect is evident in Fig.~\ref{fig:seds:comparison}, which clearly
shows that the {\sc CoStar} stellar spectra are much harder at these
energies than the other two atmosphere models.  (2) The nebular models
based on the {\sc WM-basic} SEDs show a too soft stellar spectrum
between 303 and 450 \AA\ for the models with \Teff\ $< 45\,000$~K,
which is reflected in an underprediction of the \neratio\ and \sratio\
line ratios for a given \arratio. This is also illustrated in
Fig.~\ref{fig:seds:cloudy:teff}, which shows the variation of the
\neratio\ and \arratio\ line ratios with \Teff. The \arratio\ line
ratio, which is sensitive to the number of photons able to ionize
\ion{Ar}{II} (below 450 \AA) relative to the total number of Lyman
photons, is, for a given $U$, practically independent of the stellar
model used in the photoionization model. However, large discrepancies
appear in the case of the \neratio\ line ratio.  This is because the
range of energies probed by \neratio\ depends greatly on the way the
photosphere and wind parameters of the stellar atmosphere are
defined. Hence, the large differences between the models. In
particular, the nebular models based on the {\sc WM-basic} SEDs
predict a lower \neratio\ than the other sets of models for \Teff\ $<
45\,000$~K.
(3) In the case of the nebular models based on the {\sc cmfgen} SEDs,
the discrepancies with the observations are smaller than for the other
two model grids, thus implying that the stellar spectrum between 303
and 450~\AA\ is better described by {\sc cmfgen}.

\begin{figure}[!t]
   \centering
   \includegraphics[width=8.5cm]{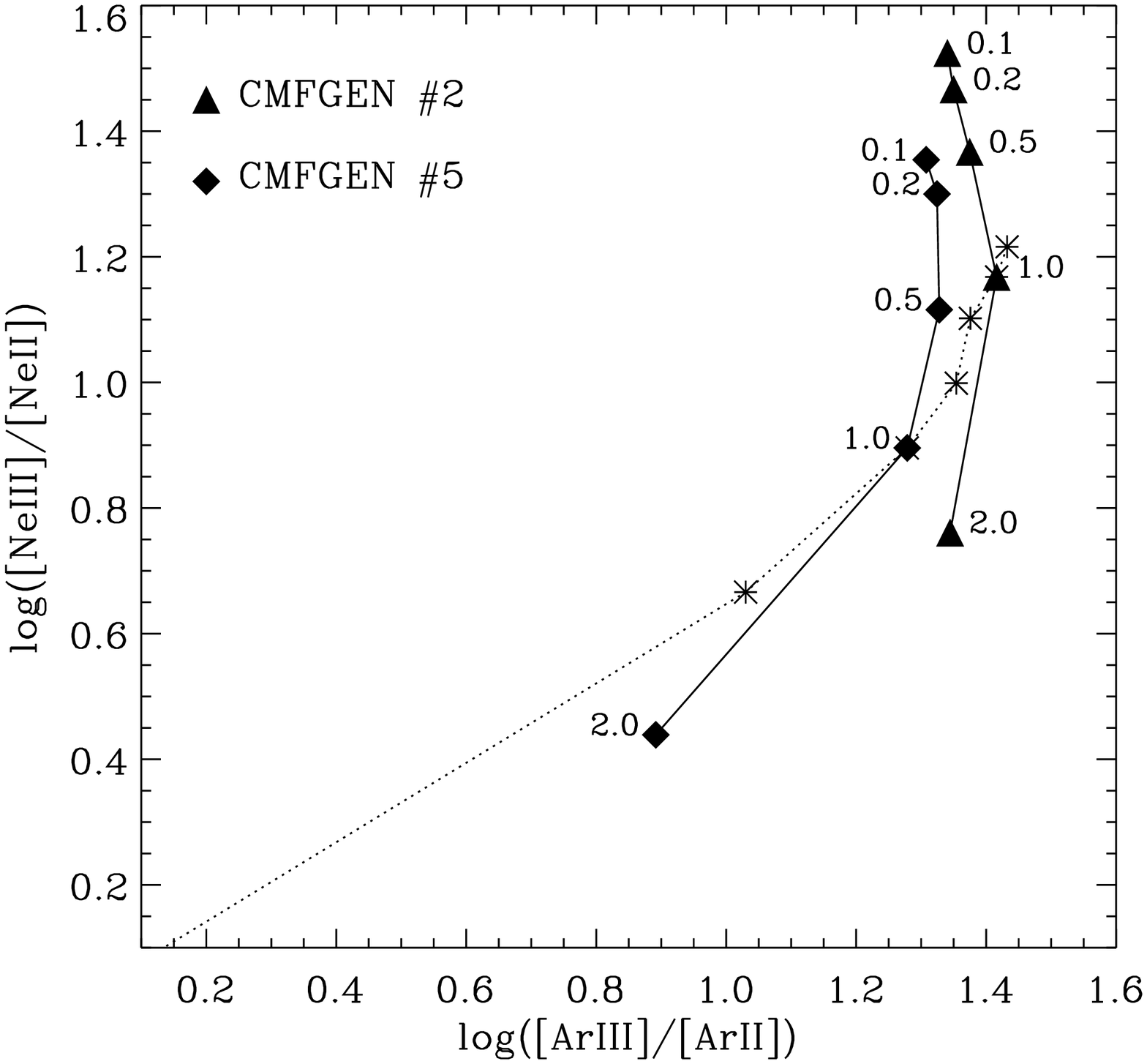}
      \caption{Variation of the nebular fine-structure line strength ratios
	\arratio~9.0/7.0 and \neratio~15.5/12.8 \mum\ as a consequence
	of modifying the stellar and nebular metallicity from 2 to
	1/10\,$Z_{\sun}$.  The adopted SEDs are models \#2 (solid
	triangles) and \#5 (solid diamonds). The dotted line connects
	the models at solar metallicity (star symbols,
	cf. Fig.~\ref{fig:seds:cloudy}). A unique ionization
	parameter, $U$, is adopted for all the models. We stress that
	unless $U$ is constrained, these line ratios cannot be used to
	derive an absolute value for \Teff.}
      \label{fig:seds:cloudy:z}
   \end{figure}

Finally, we quantify how variations in metallicity modify the
ionization structure of \HII\ regions and hence, the interpretation of
the above fine-structure line ratios in terms of the stellar content
of the nebula. Observationally, the \neratio\ line ratio observed in
\HII\ regions show a loose correlation with nebular metallicity
\citep{martin:metal}. Figure~\ref{fig:seds:cloudy:z} shows the effect
of varying the metallicity for models \#2 and \#5 on the \arratio\ and
\neratio\ line strength ratios. In Sect.~\ref{sect:seds:z} it was
discussed that, for a given effective temperature, the stellar
spectrum softens with increasing metallicity. This effect has a clear
impact on the degree of ionization of the \HII\ region: the
fine-structure line ratios decrease with increasing metallicity. The
biggest effect occurs for the \neratio\ ratio, which gets reduced by
about a factor of 10 when the metallicity increases from 1/10 to
2\,Z$_{\sun}$. Hence, the effect of metallicity is such that, for
instance, the ionization structure as traced by \neratio\ of a half
solar metallicity nebula ionized by a star with \Teff\ = $41\,010$~K
is, for the same ionization parameter, almost identical to that of a
solar metallicity nebula with a central star of \Teff\ = $48\,670$~K
conform Fig.~\ref{fig:seds:cloudy:z} \citep[also see][]{morisset03}.

Infrared line ratios have been used to infer the stellar content of
\HII\ regions \citep[e.g.][]{takahashi00,okamoto01,morisset:paperiii},
and to constrain the ages in starbursts
\citep[e.g.][]{crowther99,thornley00,spoon00}.  However, the large
dependence of these fine-structure line ratios on metallicity make
them unsuitable for such studies, unless the metallicity is known, and
moreover, its influence on the stellar energy distribution is properly
accounted for.

\section{Conclusions}
\label{sect:seds:conclusions}

We have presented a grid of eight main-sequence (dwarf) star models
ranging in effective temperature from $33\,000$ up to $51\,000$~K.  The
models were constructed using the {\sc cmfgen} code of
\cite{hillier98}. In order to investigate the variations of the
spectral appearance on metallicity, and hence, their implications for
spectral classification and the ionization structure of \HII\ regions,
we have calculated additional sets of models in which we varied the
metal content from 2 to 1/10\,$Z_{\sun}$. The main conclusions of this
study are:

\begin{itemize}

\item[$\bullet$]
  The total number of Lyman photons emitted is found to be almost
  independent of line blanketing effects and metallicity for a given
  effective temperature. This is because the flux that is blocked by
  the forest of metal lines at $\lambda < 600$ \AA\ is redistributed
  mainly within the Lyman continuum. Though some flux is removed from
  the ionizing continuum, the softening of the spectrum helps to
  conserve the number of ionizing photons.

\item[$\bullet$]
  We investigate the influence of metallicity, microturbulent velocity
  and gravity $g$ on the optical lines used to spectral type O
  stars: \HeI~$\lambda$4471 and \HeII~$\lambda$4542. We find that
  spectral type, as defined by the ratio of the equivalent widths of
  these lines, is not only a function of the effective temperature of
  the star and its luminosity class, but it depends also, albeit
  to a lower degree, on the microturbulent velocity of the stellar
  atmosphere, on metallicity and, within the luminosity class of
  dwarfs, on gravity. A change in $v_{\rm turb}$ from 5 to
  20~\kms\ can shift the spectral type by up to half a subclass (to a
  later type). A change in metallicity from 1/10 to 2 times the solar
  metallicity can shift the spectral type up to a subclass (to an
  earlier type). A decrease in $\log g$ of 0.3 dex implies a
  shift of about one spectral type (to an earlier type).

\item[$\bullet$]
  We confirm the decrease in \Teff\ for a given spectral type due to
  the inclusion of line blanketing. In particular, compared to the
  calibration of \cite{vacca96}, which is based on plane-parallel,
  pure hydrogen and helium models, we find a decrease which varies
  between $\sim 2500$ and $\sim 3800$~K. The decrease is the largest
  for stars in the spectral range of O5 to O9.

\item[$\bullet$]
  The comparison with the stellar features observed in the K-band show
  that variations in the microturbulent velocity do not have a
  significant impact on the equivalent width of the Br$\gamma$, \HeI\
  and \HeII\ lines. In the case of very hot stars variations in the
  metallicity have an appreciable influence on the equivalent width of
  the \HeII\ line. In comparing the observed \kCIV\ emission features,
  important in the spectral type calibration using this wavelength
  regime, the models show a discrepant behaviour. \kCIV\ is predicted
  to be in absorbtion while it is observed to be in emission.

\item[$\bullet$] 
  The spectral energy distribution between 303 and 450~\AA\ is best
  described by the {\sc cmfgen} SEDs.

\item[$\bullet$]
  The spectral energy distribution below $\sim 450$ \AA\ is shown to
  be highly dependent on metallicity. This is reflected by the
  behaviour of the nebular fine-structure line ratios such us
  \neratio\ 15.5/12.8 and \arratio\ 9.0/7.0 \mum. The dependence of
  these line ratios on metallicity complicates their use as diagnostic
  tools for the effective temperature of the ionizing stars in \HII\
  regions and the age dating of starburst regions in galaxies.
 
\end{itemize}

\begin{acknowledgements}
We thank the referee Rolf Kudritzki for his critical reading and
constructive comments, in particular with respect to the optical
spectral classification, which have improved this paper substantially.
We are also grateful to John Hillier for his assistance in setting up
the model grid. MRM acknowledges financial support from the NWO
Council for Physical Sciences. Model calculations were carried out
using the Beowulf cluster of the University of Amsterdam. MICE is
supported at MPE by DLR (DARA) under grants 50 QI 86108 and 50 QI
94023.
\end{acknowledgements}

\bibliographystyle{aa}
\bibliography{biblio}

\end{document}

%% file: table1.tex
\begin{table*}[!t]
\caption{Stellar and wind parameters of the model grid stars at solar metallicity.}
  \label{table:seds:grid}
  \begin{center}
    \leavevmode
    \small
    \begin{tabular}[h]{cccrcccccccc}
      \hline \hline \\[-7pt]
    \multicolumn{1}{c}{Model} &
    \multicolumn{1}{c}{\Teff} &
    \multicolumn{1}{c}{${ M_{\star} \over M_{\sun} }$} &
    \multicolumn{1}{c}{${ R_{\star} \over R_{\sun} }$} &
    \multicolumn{1}{c}{log$g$} &
    \multicolumn{1}{c}{log${ L_{\star} \over L_{\sun} }$} &
    \multicolumn{1}{c}{$H_\star$} &
    \multicolumn{1}{c}{$\log \dot{M}$} &
    \multicolumn{1}{c}{$\beta$} &
    \multicolumn{1}{c}{$V_\infty$} &
    \multicolumn{1}{c}{$\log Q_0$} &
    \multicolumn{1}{c}{$\log Q_1$} \\

    \multicolumn{1}{c}{} &
    \multicolumn{1}{c}{(K)} &
    \multicolumn{1}{c}{} &
    \multicolumn{1}{c}{} &
    \multicolumn{1}{c}{(cm\,s$^{-2}$)} &
    \multicolumn{1}{c}{} &
    \multicolumn{1}{c}{($10^{-4}R_{\star}$)} &
    \multicolumn{1}{c}{($M_{\sun}$/yr)} &
    \multicolumn{1}{c}{} &
    \multicolumn{1}{c}{(\kms)} &
    \multicolumn{2}{c}{(photons~s$^{-1}$)} \\[5pt] \hline \\[-7pt]

1 & 51230 & 87.6 & 13.2 & 4.139 & 6.032 & 7.703 & $-5.209$ & 0.8 & 3450
& 49.87 & 49.38\\
2 & 48670 & 68.9 & 12.3 & 4.096 & 5.882 & 8.308 & $-5.375$ & 0.8 & 3240
& 49.69 & 49.18\\
3 & 46120 & 56.6 & 11.4 & 4.077 & 5.772 & 8.375 & $-5.599$ & 0.8 & 3140
& 49.50 & 48.95\\
4 & 43560 & 45.2 & 10.7 & 4.034 & 5.568 & 8.971 & $-5.805$ & 0.8 & 2950
& 49.31 & 48.72\\
5 & 41010 & 37.7 & 10.0 & 4.014 & 5.404 & 9.046 & $-6.072$ & 0.8 & 2850
& 49.08 & 48.44\\
6 & 38450 & 30.8 &  9.3 & 3.989 & 5.229 & 9.314 & $-6.369$ & 0.8 & 2720
& 48.82 & 48.05\\
7 & 35900 & 25.4 &  8.8 & 3.954 & 5.062 & 9.712 & $-6.674$ & 0.8 & 2570
& 48.51 & 47.43\\
8 & 33340 & 21.2 &  8.3 & 3.926 & 4.883 & 9.934 & $-7.038$ & 0.8 & 2450
& 48.06 & 46.07\\[3pt]

\hline

    \end{tabular}
  \end{center}
\end{table*} 

%% file: table2.tex
\begin{table}[!t]
\caption{Chemical composition adopted for the models with
$Z=Z_{\sun}$.}
  \label{table:seds:chemical}
  \begin{center}
    \leavevmode
    \small
    \begin{tabular}[h]{cc}
      \hline \hline \\[-7pt]
      Element & Mass fraction \\[5pt] \hline \\[-7pt]
    H  & 0.7023		      \\
    He & 0.2820		      \\
    C  & $3.050\times10^{-3}$ \\
    N  & $1.100\times10^{-3}$ \\
    O  & $9.540\times10^{-3}$ \\
    Si & $6.990\times10^{-4}$ \\
    Fe & $1.360\times10^{-3}$ \\[3pt]
    \hline
    \end{tabular}
  \end{center}
\end{table} 


%% file: table3.tex
\begin{table}[!t]
\caption{Variation of the scaleheight and the mass loss rate with
metallicity for models \# 2 and \#5.}
  \label{table:seds:grid:z}
  \begin{center}
    \leavevmode
    \small
    \begin{tabular}[h]{cccccc}
      \hline \hline \\[-7pt]  

    \multicolumn{1}{c}{} & 
    \multicolumn{2}{c}{Model \#2} & &  
    \multicolumn{2}{c}{Model \#5} \\  \cline{2-3} \cline{5-6} \\[-5pt]

    \multicolumn{1}{c}{} &
    \multicolumn{1}{c}{$H_\star$} &
    \multicolumn{1}{c}{$\log \dot{M}$} &
    \multicolumn{1}{c}{} &
    \multicolumn{1}{c}{$H_\star$} &
    \multicolumn{1}{c}{$\log \dot{M}$} \\

    \multicolumn{1}{c}{Z/Z$_{\sun}$} &
    \multicolumn{1}{c}{($10^{-4}R_{\star}$)} &
    \multicolumn{1}{c}{($M_{\sun}$/yr)} &
    \multicolumn{1}{c}{} &
    \multicolumn{1}{c}{($10^{-4}R_{\star}$)} &
    \multicolumn{1}{c}{($M_{\sun}$/yr)} \\[5pt] \hline \\[-7pt]  

2    & 7.817 & $-5.120$ && 8.511 & $-5.815$\\
1/2  & 8.554 & $-5.631$ && 9.314 & $-6.328$\\
1/5  & 8.702 & $-5.971$ && 9.474 & $-6.665$\\
1/10 & 8.751 & $-6.225$ && 9.528 & $-6.921$\\[3pt]

\hline

    \end{tabular}
  \end{center}
\end{table}


%% file: table4.tex
   \begin{table*}[!t]
     \caption{Schematic comparison of assumptions made in the codes
        used for the EUV comparion. CMF implies a co-moving frame
        treatment of line transfer; SE denotes that the state of the
        gas is derived from statistical equilibrium (i.e. non-LTE). 
        See text for a discussion.}
     \label{table:seds:codes}
      \begin{center}
        \leavevmode
        \small
       \begin{tabular}{llll}
         \hline \hline \\[-7pt] 

         Treatment                  &   {\sc cmfgen} &  {\sc WM-basic}  &  {\sc CoStar} \\[5pt] \hline \\[-7pt]

         {\it Temperature structure}&  Consistent    &  Consistent      &  Grey-LTE with blanketing feedback\\
         {\it Density structure}    &                &                  &                            \\
         \hspace{3mm} Photosphere   &  Constant $H$  &  Consistent      &  Grey-LTE + $p_{R}$-cont   \\
         \hspace{3mm} Wind          &  $\beta$-law   &  Consistent      &  $\beta$-law               \\
         {\it Non-LTE treatment}    &  Consistent    &  Consistent      &  Consistent (except Fe)    \\
         \hspace{3mm} Line Transfer &  CMF           &  CMF             &  Sobolev                   \\
         \hspace{3mm} Iron-group    &  SE            &  SE              &  Modified Nebular          \\  
         {\it Number of transitions/lines}      &                &                  &                            \\
         \hspace{3mm} In SE         &  $\sim$9\,000  &  $\sim$30\,000   &  $\sim$1\,000              \\
         \hspace{3mm} In spectrum   &  $\sim$17\,000 & $\sim$4\,000\,000&  $\sim$100\,000            \\
         {\it Blanketing}           &  Consistent    &  Consistent      &  Blanketing factors        \\[3pt]
         \hline

       \end{tabular}
      \end{center}
   \end{table*}


%% file: table5.tex
\begin{table}[!t]
\caption{Predicted number of ionizing photons $Q_0$ and $Q_1$ for
   \# 2 and \#5 as a function of metal abundance.}
  \label{table:seds:q:z}
  \begin{center}
    \leavevmode
    \small
    \begin{tabular}[h]{cccccc}
      \hline \hline \\[-7pt]  

    \multicolumn{1}{c}{} &
    \multicolumn{2}{c}{Model \#2} & &  
    \multicolumn{2}{c}{Model \#5} \\  \cline{2-3} \cline{5-6} \\[-5pt]

    \multicolumn{1}{c}{$Z/Z_{\sun}$} &
    \multicolumn{1}{c}{log$Q_0$} &
    \multicolumn{1}{c}{log$Q_1$} &
    \multicolumn{1}{c}{} &
    \multicolumn{1}{c}{log$Q_0$} &
    \multicolumn{1}{c}{log$Q_1$} \\ [5pt] \hline \\[-7pt]  

2    & 49.70 &  49.16  && 49.09 & 48.40  \\
1    & 49.69 &  49.18  && 49.08 & 48.44  \\
1/2  & 49.69 &  49.19  && 49.07 & 48.46  \\
1/5  & 49.68 &  49.20  && 49.07 & 48.47  \\
1/10 & 49.67 &  49.21  && 49.06 & 48.48  \\[3pt] \hline 

    \end{tabular}
  \end{center}
\end{table}
